\documentclass[journal]{IEEEtran}

\usepackage{epsf,psfrag,amssymb,amsfonts,color,cite,fancybox}
\usepackage[mathscr]{eucal}
\usepackage[dvips]{graphicx}
\usepackage{caption}
\usepackage{subcaption}
\usepackage{ifpdf}
\usepackage{longtable}	
\usepackage{multicol}
\usepackage{float}
\usepackage{epstopdf}
\usepackage{bm}
\usepackage{multirow}
\usepackage[scientific-notation=true]{siunitx}

\ifCLASSINFOpdf

\else

\fi

\usepackage[cmex10]{amsmath}
\usepackage[caption=false,font=footnotesize]{subfig}

\begin{document}

%\title{Towards a PMU-Based Approach to Diagnose Various Oscillations in Power Systems}

\title{PMU-Based Estimation of Dynamic State Jacobian Matrix}%{Hybrid Approach to Estimate Real-Time Dynamic State Jacobian Matrix in Stochastic Power System}

\author{Xiaozhe Wang,~\IEEEmembership{Member,~IEEE,}
        Konstantin Turitsyn,~\IEEEmembership{Member,~IEEE.}
        % <-this % stops a space
%\thanks{This work was supported by the Consortium for Electric Reliability
%Technology Solutions provided by U.S. Department No. DE-FC26-
%09NT43321.}
\thanks{Xiaozhe Wang and Konstantin Turitsyn are with the Department of Mechanical Engineering, MIT, Cambridge, MA, 02139. email: xw264@cornell.edu, turitsyn@mit.edu.}% <-this % stops a space
}

\maketitle

\begin{abstract}
In this paper, a hybrid measurement and model-based method is proposed which can estimate the dynamic state Jacobian matrix in near real-time. The proposed method is computationally efficient and robust to the variation of network topology. Since the estimated Jacobian matrix carries significant information on system dynamics and states, it can be utilized in various applications. In particular, two application of the estimated Jacobian matrix in online oscillation analysis, stability monitoring and control are illustrated with numerical examples. In addition, a side-product of the proposed method can facilitate model validation by approximating the damping of generators.
\end{abstract}


\IEEEpeerreviewmaketitle

\section{Introduction}
The main goal of the power system operator is to maintain the system in the normal secure state despite varying operating conditions. Achievement of this goal necessitates continuous monitoring of the system condition, identification of the operating state and determination of the necessary control actions, which are referred to as the security analysis of a system \cite{Abur:book}. %For example, the N-1 contingency analysis is performed periodically by energy management system (EMS) to help the power system operator determine whether the system can withstand a set of credible contingencies given the current operating condition, and further decide whether corrective control such as generation re-dispatch is required \cite{Chiang:book}.
Security assessment such as contingency analysis, optimal power flow and generation re-dispatch typically involves repeated computation based on the full nonlinear power system model, leading to enormous computational burden \cite{Chiang:book}. In addition, the conventional security analysis strongly relies on an accurate system model and network topology, which may not be available due to erroneous telemetry from remotely monitored circuit breaker \cite{Chen:2014}.
%Moreover, the constantly increasing penetration of renewable energy as well as load variation introduce stochastic dynamics to the system which may affect the security of the system and make the conventional methods for deterministic systems inaccurate \cite{crow:2012}\cite{Hines:2015}.
The above factors pose great challenges to the static and dynamic security analysis for the modern power grids and make the online security analysis even more formidable. Wide adoption of phasor measurement units (PMUs) greatly facilitates online monitoring of system states and its dynamic behaviors, offering unique opportunities for the development of measurement-based security analysis methods in power system \cite{Wangxz:2015}-\cite{Zhou:2009PES}.

In this paper, we propose a hybrid measurement and model-based method to estimate the dynamic state Jacobian matrix in near real-time, which provides invaluable information on system states and conditions. The proposed hybrid method nicely combines the statistical properties extracted from the PMU measurements and the inherent generator physics, working as a grey box bridging the measurement and the model. In addition, the estimation of the Jacobian matrix does not require exhaustive computational effort and can be done in near real-time. Furthermore, the estimation method is robust to network topology and parameter changes.
Conventionally, the Jacobian matrix can be constructed based on state estimation results provided that an accurate dynamic model and network parameter values are available \cite{Abur:book}. However, as discussed before, up-to-date network topology and parameters may not be available, so the conventional method may give rise to imprecise estimations. In contrast, the proposed hybrid method does not depend on the network parameter values, and provides a robust alternative to traditional state estimation-based approaches in situations where uncertainty in parameters and system topology is an issue.

%The last but not the least, the proposed method incorporates the stochastic dynamics inside a power system and thus is more applicable to power systems with high penetration of renewable energies or large load variation compared to deterministic model-based methods.

The estimated dynamic state Jacobian matrix can be utilized in various applications. First, it will be shown that the estimated Jacobian matrix can facilitate online oscillation analysis and control.
The conventional model-based method for oscillation analysis relies on an accurate power system model and requires exhaustive computational effort \cite{Zhou:2007}, which is not suitable for online oscillation monitoring. Hence, much attention has been put for measurement-based methods which are deemed to be fast and robust to parameter change.
Particularly, measurement-based mode shape estimation methods have been proposed by using Prony analysis \cite{Scharf:1990}, spectral methods \cite{Trudnowski:2008}\cite{Zhou:2013PES}, frequency domain decomposition \cite{Venkatasubramanian:2008}, subspace method \cite{Zhou:2007}\cite{Liu:2011PES}, transfer function method \cite{Zhou:2009PES}, etc. Unlike these previous studies, online oscillation analysis and control technique developed in this paper relies on online estimation of the Jacobian matrix. Hence, participation factor of the excited model is achievable, which conversely can not be extracted from the purely measurement-based approaches. %In addition to the oscillation mode shape, frequency, and damping ratio, it provides an
As the participation factor provides relative contribution of each state in the corresponding mode, it can naturally be used for locating oscillation source and designing control actions \cite{Kundur:book}. A numerical example will be presented to show how the estimated Jacobian matrix may help online oscillation analysis and control. A relevant work \cite{Dobson:2015} applies a formula to damp interarea oscillations by generator re-dispatch based on eigenvalue sensitivity instead of participation factor.
%As the estimate of the participation factor depends on left eigenvector, it can not be extracted from the purely measurement-based approaches.
%In this paper,

Another application of the estimated Jacobian matrix is online steady-state stability monitoring and control. Measurement-based approaches for real-time rotor angle and short-term voltage stability monitoring have been explored in many previous works (e.g. \cite{CCLiu:2011}\cite{Ajjarapu:2013}). Regarding steady-state stability, the authors of \cite{Hines:2015}\cite{Hines:2014} have proposed statistical stability indicators for stochastic power systems based on the phenomenon of critical slowing down. In this paper, the proposed stability indicators are based on the online estimate of the Jacobian matrix and hence the state matrix. Firstly, the critical eigenvalue (i.e. the eigenvalue with maximum real part) of the state matrix can be used as a good measure of proximity to instability \cite{Kundur:1992}. In addition to a simple stability indicator, the left and right eigenvectors of the critical eigenvalue can be utilized to predict the response of the system and design emergency control measures \cite{Hill:1993}\cite{Cutsem:book}. A numerical example will be given to illustrate how the estimated Jacobian matrix by the proposed method may contribute to online stability monitoring and control.

Apart from the above applications, we believe that the estimated Jacobian matrix can also facilitate other emergency decision-making in power system operation such as generation re-dispatch and congestion relief, which requires further investigation. Besides the estimated Jacobian matrix, a side-product of the method provides an approach to estimate damping of the synchronous generators. A numerical example will be given to illustrate this application in model validation.

The rest of the paper is organized as follows. Section \ref{sectionmodel} briefly introduces the power system dynamic model and then elaborates the proposed hybrid measurement model-based method. Section \ref{sectionapplications} presents the applications of the estimated Jacobian matrix in online oscillation analysis and stability monitoring via numerical examples. The application of the derived relation in model validation will also be presented. Conclusions and perspectives are given in Section \ref{sectionconclusion}.

\section{hybrid approach to estimate jacobian matrix}\label{sectionmodel}
We consider the general power system dynamic model: %with stochastic generation and load:
\begin{eqnarray}
\dot{\bm{x}}&=&\bm{f}({\bm{x},\bm{y}})\label{fast ode}\\%+\overline{\Sigma}\bm{\bar{\xi}}\\
\bm{0}&=&\bm{g}({\bm{x},\bm{y}})\label{algebraic eqn}
\end{eqnarray}
Equation (\ref{fast ode}) describes dynamics of generators, %and their associated control as well as load dynamics,
and Eqn (\ref{algebraic eqn}) describes the electrical transmission system and the internal static behaviors of passive devices. $f$ and $g$ are continuous functions, vectors $\bm{x}\in\mathbb{R}^{n_{\bm{x}}}$ and $\bm{y}\in\mathbb{R}^{n_{\bm{y}}}$ are the corresponding state variables (generator rotor angles, rotor speeds) and algebraic variables (bus voltages, bus angles) \cite{Wangxz:CAS}.

In this paper, we focus on ambient oscillations around stable steady state, dominated mainly by the dynamics of generator angles. Hence, we demonstrate the proposed technique using the classical generator model, which can be regarded as an equivalent generator model of aggregated generators \cite{Chow:2014}. In Appendix \ref{higherordermodel}, the results are numerically validated on higher-order models.
We assume that the mechanical power injection of each generator is experiencing standard Gaussian noise possibly due to renewables, i.e., the mechanical power for Generator $i$ is $P_{m_i}+\sigma_i\xi_i$, where $\xi_i$ is a standard Gaussian noise, and $\sigma_i^2$ is the noise variance.
Suppose $g_{\bm{y}}$ is nonsingular, (\ref{fast ode})-(\ref{algebraic eqn}) can be represented as:
\begin{eqnarray}
\dot{\bm{\delta}}&=&\bm{\omega}\label{swing-1}\\
M\dot{\bm{\omega}}&=&\bm{P_m}-\bm{P_e}-{D}\bm{\omega}+{\Sigma}\bm{\xi}\label{swing-2}
\end{eqnarray}
with algebraic variables $\bm{y}$ eliminated. Particularly, $\bm{\delta}=[\delta_1,...\delta_n]^T$ is a vector of generator rotor angles, $\bm{\omega}=[\omega_1,...\omega_n]^T$ is a vector of generator rotor speeds, $M=\mbox{diag}(M_1,...M_n)$ whose diagonal entries are the moment inertia constants, $D=\mbox{diag}(D_1,...D_n)$ whose diagonal entries are damping factors, $\bm{P_m}=[P_{m_1},...P_{m_n}]^T$ is a vector of generators' mechanical power, $\bm{P_e}=[P_{e_1},...P_{e_2}]^T$ is a vector of generators' electrical power. In addition, $\bm{{\xi}}$ is a vector of independent standard Gaussian random variables representing the variation of power injections, and $\Sigma=\mbox{diag}(\sigma_1,...\sigma_n)$ is the covariance matrix.
For the sake of simplicity, in this work we model the loads as constant impedances. In the future, more realistic models can be incorporated in a similar way as higher-order generator models briefly discussed in Appendix \ref{higherordermodel}.
%{\color{red} Loads ?}
%Within this model the loads can be modeled either as constant impedances that are directly introduced

%The structure-preserving extension of the classical model allows for natural incorporation of loads, that are modeled as composition of fixed impedance and frequency dependent power loads \cite{StructurePreserving}.

Linearizing (\ref{swing-1})-(\ref{swing-2}) gives the following:
\begin{eqnarray}
\left[\begin{array}{c}\dot{\bm{\delta}}\\\dot{\bm{\omega}}\end{array}\right]&=&
\left[\begin{array}{cc}{{0}}&{I_n}\\-M^{-1}\frac{\partial{\bm{P_e}}}{\partial{\bm{\delta}}}&-M^{-1}D\end{array}\right]
\left[\begin{array}{c}{\bm{\delta}}\\{\bm{\omega}}\end{array}\right]\nonumber\\
&&+\left[\begin{array}{c}0\\M^{-1}\Sigma\end{array}\right]\bm{\xi}\label{swing-matrix}
\end{eqnarray}
Let $\bm{x}=[\bm{\delta},\bm{\omega}]^T$, $A=\left[\begin{array}{cc}{{0}}&{I_n}\\-M^{-1}\frac{\partial{\bm{P_e}}}{\partial{\bm{\delta}}}&-M^{-1}D\end{array}\right]$, $B=[0,M^{-1}\Sigma]^T$, then (\ref{swing-matrix}) takes the form:
\begin{equation}
\dot{\bm{x}}=A\bm{x}+B\bm{\xi}
\end{equation}
Specifically, if state matrix $A$ is stable, the stationary covariance matrix $C_{\bm{x}\bm{x}}=\left[\begin{array}{cc}C_{\bm{\delta}{\bm{\delta}}}&C_{\bm{\delta}{\bm{\omega}}}\\C_{\bm{\omega}{\bm{\delta}}}&C_{\bm{\omega}{\bm{\omega}}}\end{array}\right]$ can be shown to satisfy the following Lyapunov equation \cite{Hines:2015}\cite{Gardiner:2009}:
\begin{equation}
AC_{\bm{x}\bm{x}}+C_{\bm{x}\bm{x}}A^T=-BB^T \label{lyapunov}
\end{equation}
which nicely combine the statistical properties of states and the model knowledge.
This relation between the covariance matrix $C_{\bm{x}\bm{x}}$ and the system state matrix is important and has been utilized in \cite{Hines:2015} to compute the covariance matrix $C_{\bm{x}\bm{x}}$ based on the model knowledge $A$ and $B$. In this paper, we utilize this relation the other way round, which is to approximate the state matrix $A$ according to the measurements $C_{\bm{x}\bm{x}}$.

Substituting the detailed expressions of $A$ and $B$ to (\ref{lyapunov}) and conducting algebraic simplification, we obtain that:
\begin{eqnarray}
C_{\bm{\delta}{\bm{\omega}}}&=&0\\
C_{\bm{\delta}{\bm{\delta}}}&=&(\frac{\partial{\bm{P_e}}}{\partial{\bm{\delta}}})^{-1}MC_{\bm{\omega}{\bm{\omega}}}\label{rdd}\\
C_{\bm{\omega}{\bm{\omega}}}&=&\frac{1}{2}M^{-1}D^{-1}\Sigma^2\label{rww}
\end{eqnarray}
The relations given by (\ref{rdd})-(\ref{rww}) link the measurements of stochastic variations to the generator physical model, which provides an ingenious way to estimate system dynamic state Jacobian matrix or parameter values from the measurements. Firstly, given that the parameter values of $M$ are available, and $C_{\bm{\delta}\bm{\delta}}$, $C_{\bm{\omega}{\bm{\omega}}}$ can be acquired through the PMU measurements, the Jacobian matrix $\frac{\partial{\bm{P_e}}}{\partial{\bm{\delta}}}$ can be obtained from (\ref{rdd}), and furthermore the system state matrix $A$ can be readily constructed. Secondly, (\ref{rww}) can be used to estimate the parameter values of $D$ in dynamic equivalencing provided that $M$ and $\Sigma$ are available and $C_{\bm{\omega}{\bm{\omega}}}$ can be obtained from the PMU measurements. The applications of the derived relations will be detailed elaborated in Section \ref{sectionapplications}. In the rest of this Section, we will use a numerical example to illustrate the validity of the proposed method.

\subsection{Numerical Illustration}

We consider the standard WSCC 3-generator, 9-bus system model (see, e.g. \cite{Chiang:book}). %The parameter values of the system are available online:...
The system model in the center-of-inertia (COI) formulation is presented as below:
\begin{eqnarray}
\dot{\tilde{\delta}}_1&=&\tilde{\omega}_1\label{9bus-1}\\
\dot{\tilde{\delta}}_2&=&\tilde{\omega}_2\\
M_1\dot{\tilde{\omega}}_1&=&P_{m_1}-P_{e_1}-\frac{M_1}{M_T}P_{coi}-D_1\tilde{\omega}_1+\sigma_1\xi_1\\
M_2\dot{\tilde{\omega}}_2&=&P_{m_2}-P_{e_2}-\frac{M_2}{M_T}P_{coi}-D_2\tilde{\omega}_2+\sigma_2\xi_2\label{9bus-2}
\end{eqnarray}
where $\delta_0=\frac{1}{M_T}\sum_{i=1}^{3}M_i\delta_i$, $\omega_0=\frac{1}{M_T}\sum_{i=1}^{3}M_i\omega_i$, $M_T=\sum_{i=1}^{3}M_i$, $\tilde{\delta}_i=\delta_i-\delta_0$, $\tilde{\omega}_i=\omega_i-\omega_0$, for $i=1,2,3$,
and
\begin{eqnarray}
P_{e_i}&=&\sum_{j=1}^{3}E_iE_j(G_{ij}\cos(\tilde{\delta}_i-\tilde{\delta}_j)+B_{ij}\sin(\tilde{\delta}_i-\tilde{\delta}_j))\nonumber\\
P_{coi}&=&\sum_{i=1}^{3}(P_{m_i}-P_{e_i})
\end{eqnarray}
The parameter values in this examples are: $M_1=0.63$, $M_2=0.34$, $M_2=0.16$; $D_1=0.63$, $D_2=0.34$, $D_3=0.16$; $P_{m_1}=0.72$ p.u., $P_{m_2}=1.63$ p.u., $P_{m_3}=0.85$ p.u.; $E_1=1.057$ p.u., $E_2=1.050$ p.u., $E_3=1.017$ p.u..
Because the following relations that $\tilde{\delta}_3=-\frac{M_1\tilde{\delta}_1+M_2\tilde{\delta}_2}{M_3}$ and $\tilde{\omega}_3=-\frac{M_1\tilde{\omega}_1+M_2\tilde{\omega}_2}{M_3}$ hold in the COI formulation, $\tilde{\delta}_3$ and $\tilde{\omega}_3$ depending on the other state variables can be obtained without integration.

The system state matrix is as follows:
\begin{equation}
A=\left[\begin{array}{cc|cc}0&0&1&0\\0&0&0&1\\\hline\multicolumn{2}{c}{\multirow{2}{*}{J}}\vline\hspace{-0.002in}&-\frac{D_1}{M_1}&0\\&&0&-\frac{D_2}{M_2}\end{array}\right]\label{A}
\end{equation}
where $J=-M^{-1}(\frac{\partial\bm{P_e}}{\partial\bm{\tilde{\delta}}}+M\frac{1}{M_T}\frac{\partial P_{coi}}{\partial\bm{\tilde{\delta}}})$, for $i=1,2$. Let $(\frac{\partial\bm{P_e}}{\partial\bm{\tilde{\delta}}})_{coi}=\frac{\partial\bm{P_e}}{\partial\bm{\tilde{\delta}}}+M\frac{1}{M_T}\frac{\partial P_{coi}}{\partial\bm{\tilde{\delta}}}$, then we have
\begin{equation}
\small{((\frac{\partial\bm{P_e}}{\partial\bm{\tilde{\delta}}})_{coi})_{ij}}=\left\{\begin{array}{l}E_iE_j(G_{ij}\sin(\tilde{\delta}_i-\tilde{\delta}_j)-B_{ij}\cos(\tilde{\delta}_i-\tilde{\delta}_j))\\
+\frac{M_i}{M_T}\frac{\partial P_{coi}}{\partial\tilde{\delta_i}} \hspace{1.2in} \mbox{if $i\not=j$}\\
\sum^{n}_{k=1}E_iE_k(G_{ik}\sin(\tilde{\delta}_i-\tilde{\delta}_k)\\
+B_{ik}\cos(\tilde{\delta}_i-\tilde{\delta}_k))+\frac{M_i}{M_T}\frac{\partial P_{coi}}{\partial\tilde{\delta_i}} \hspace{0.1in} \mbox{if $i=j$}
\end{array}\right.\label{dpeddcoi}
\end{equation}
where $\frac{\partial P_{coi}}{\partial \tilde{\delta_i}}=2\sum_{k\not=i}E_iE_kG_{ik}\sin(\tilde{\delta}_i-\tilde{\delta}_k)$.

If the network parameter values as well as system states are available, the Jacobian matrix $(\frac{\partial\bm{P_e}}{\partial\bm{\tilde{\delta}}})_{coi}$ can be directly computed according to (\ref{dpeddcoi}). However, the system topology and line model parameter values are subjected to continuous perturbations. Therefore, the exact knowledge of network topology with up-to-date network parameter values may not be available. In addition, the control faults and transmission delays may also lead to imprecise knowledge of the network parameter values.

In contrast, the proposed method does not require the knowledge of network topology and network parameters. Suppose the parameter values $M$ of the generators are available, and $C_{\bm{\tilde{\omega}}{\bm{\tilde{\omega}}}}$, $C_{\bm{\tilde{\delta}}{\bm{\tilde{\delta}}}}$ can be obtained from the PMU measurement, we can estimate the Jacobian matrix by: % as long as we have the parameter values $M$ of the generators are accurate:
\begin{equation}
(\frac{\partial\bm{P_e}}{\partial\bm{\tilde{\delta}}})_{coi}^{\star}=MC_{\bm{\tilde{\omega}}{\bm{\tilde{\omega}}}}C^{-1}_{\bm{\tilde{\delta}}{\bm{\tilde{\delta}}}}\label{proposemethod}
\end{equation}
Particularly, we use $^\star$ to denote the Jacobian matrix estimated by the proposed method.

\begin{figure}[!ht]
\centering
\begin{subfigure}[t]{0.52\linewidth}
\includegraphics[width=1.8in ,keepaspectratio=true,angle=0]{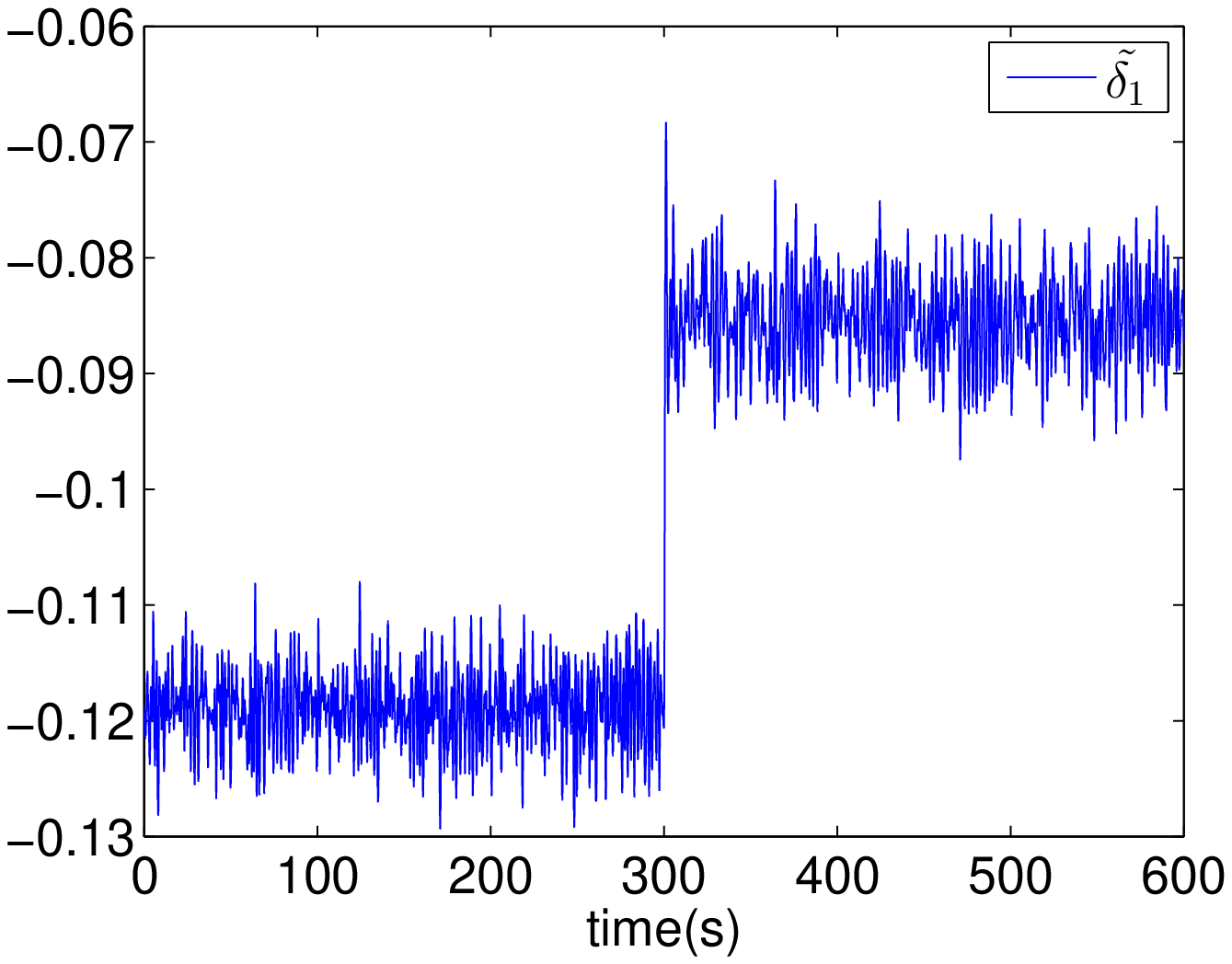}
\caption{Trajectory of $\tilde{\delta}_1$ on [0s,600s]}\label{d1-9}
\end{subfigure}%
%\begin{subfigure}[t]{0.48\linewidth}
%\includegraphics[width=1.8in ,keepaspectratio=true,angle=0]{d2_9_2.eps}
%\caption{Trajectory of $\tilde{\delta}_2$ on [0s,600s]}\label{d2-9}
%\end{subfigure}
\begin{subfigure}[t]{0.5\linewidth}
\includegraphics[width=1.8in ,keepaspectratio=true,angle=0]{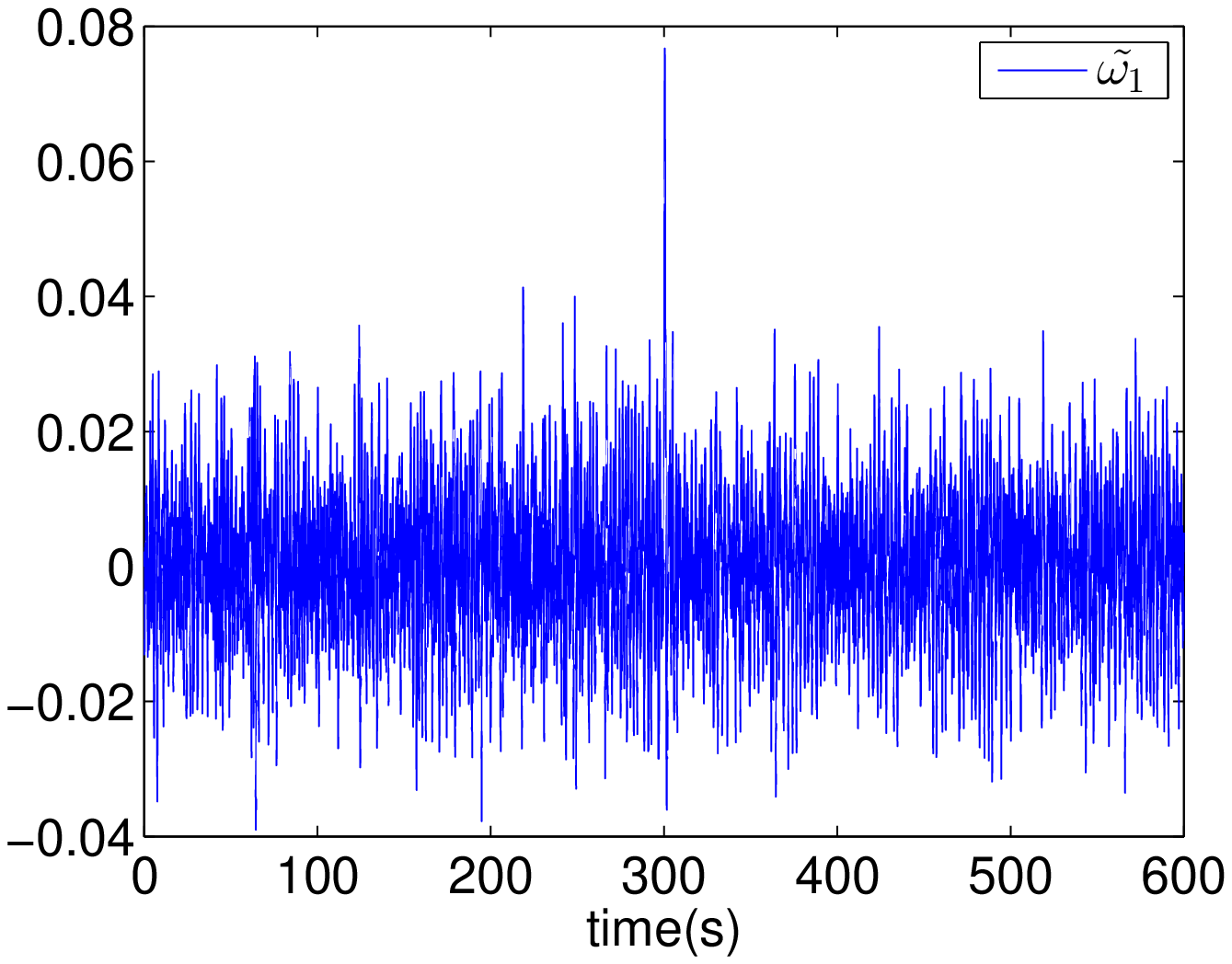}
\caption{Trajectory of $\tilde{\omega}_1$ on [0s,600s]}\label{w1-9}
\end{subfigure}
%\begin{subfigure}[t]{0.5\linewidth}
%\includegraphics[width=1.8in ,keepaspectratio=true,angle=0]{w2_9_2.eps}
%\caption{Trajectory of $\tilde{\omega}_2$ on [0s,600s]}\label{w2-9}
%\end{subfigure}
%\includegraphics[width=2.5in,keepaspectratio=true,angle=0]{WDtimeseries.eps}
\caption{Trajectories of the state variables for the 9-bus system in COI reference.}\label{9bus}
\end{figure}

In order to show that the proposed method is not affected by network topology, we conduct the following numerical experiment. Assuming that the transient reactance $x_d'$ of Generator 1 increases from $0.0608$ p.u. to $0.1824$ p.u. at 300.01s, mimicking a ling loss between the generator internal node and its terminal bus \cite{Pai:2012}. Let $\sigma_1=\sigma_2=0.01$, %and $x_d'$ increase to $0.503 $ p.u. from $0.1198$ p.u. at 100.01s,
the trajectories of some state variables in system (\ref{9bus-1})-(\ref{9bus-2})
before and after the contingency are presented in Fig. \ref{9bus}, %Specifically, Fig. \ref{d1-9-zoom} is the trajectory of $\tilde{\delta}_1$ before the contingency by zooming the corresponding part in Fig. \ref{d1-9}.
from which we see that the system is able to maintain stability after the contingency, and the state variables are always fluctuating around nominal states due to the variation of power injections.

Before the contingency and without considering stochastic perturbations, i.e., $\sigma_1=\sigma_2=0$, $(\frac{\partial\bm{P_e}}{\partial\bm{\tilde{\delta}}})_{coi}$ is a constant matrix which can be easily obtained from (\ref{dpeddcoi}):
\begin{equation}
(\frac{\partial\bm{P_e}}{\partial\bm{\tilde{\delta}}})_{coi}=\left[ \begin{array}{cc} 8.053 & 1.240\\ 2.802 & 5.085\end{array}\right]\label{det_before}
\end{equation}
We want to show that the matrix obtained from the proposed method is close to this deterministic matrix. %Next, we apply the proposed method.

Firstly, $C_{\bm{\tilde{\omega}}{\bm{\tilde{\omega}}}}$ and $C_{\bm{\tilde{\delta}}{\bm{\tilde{\delta}}}}$ before the contingency can be calculated based on the system trajectories on [0s, 300s]:
\begin{eqnarray}
C_{\bm{\tilde{\delta}}{\bm{\tilde{\delta}}}}&=&10^{-4}\times\left[ \begin{array}{cc}  0.106 & -0.0483\\ -0.0483 &0.359\end{array}\right]\nonumber\\
C_{\bm{\tilde{\omega}}{\bm{\tilde{\omega}}}}&=&10^{-3}\times\left[ \begin{array}{cc}  0.123 & 0.008\\ 0.008 &0.514\end{array}\right]\nonumber
\end{eqnarray}
and therefore $(\frac{\partial\bm{P_e}}{\partial\bm{\tilde{\delta}}})_{coi}$ can be computed by the proposed hybrid method according to (\ref{proposemethod}):
\begin{equation}
(\frac{\partial\bm{P_e}}{\partial\bm{\tilde{\delta}}})_{coi}^{\star}=\left[ \begin{array}{cc} 7.806 & 1.192\\  2.642 &  5.214 \end{array}\right]
\end{equation}
It is observed that $(\frac{\partial\bm{P_e}}{\partial\bm{\tilde{\delta}}})_{coi}^{\star}$ and $(\frac{\partial\bm{P_e}}{\partial\bm{\tilde{\delta}}})_{coi}$ are close to each other. Specifically, the estimation error is:
\begin{equation}
\frac{\|(\frac{\partial\bm{P_e}}{\partial\bm{\tilde{\delta}}})_{coi}^\star-(\frac{\partial\bm{P_e}}{\partial\bm{\tilde{\delta}}})_{coi}\|_F}{\|(\frac{\partial\bm{P_e}}{\partial\bm{\tilde{\delta}}})_{coi}\|_F}=3.25\%\label{matrixdistance} \end{equation}
where $\|\|_F$ denotes the Frobenius norm of a matrix measuring the distance between two matrixes. Equation (\ref{matrixdistance}) demonstrates the validity and accuracy of the proposed method.
%However, we can not claim that which one is more accurate and trustworthy.

To highlight the value of the proposed hybrid method, we assume that the change of $x_d'$ of Generator 1 is undetected. Therefore the Jacobian matrix $(\frac{\partial\bm{P_e}}{\partial\bm{\tilde{\delta}}})_{coi}$ after the contingency is regarded to be the same as shown in (\ref{det_before}). In contrast, the proposed hybrid method gives:
\begin{eqnarray}
\overline{C_{\bm{\tilde{\delta}}{\bm{\tilde{\delta}}}}}&=&10^{-4}\times\left[ \begin{array}{cc}  0.139 & -0.108\\ -0.108 &0.495\end{array}\right]\nonumber\\
\overline{C_{\bm{\tilde{\omega}}{\bm{\tilde{\omega}}}}}&=&10^{-3}\times\left[ \begin{array}{cc}  0.113 & -0.024\\ -0.024 &0.658\end{array}\right]\nonumber\\
\overline{(\frac{\partial\bm{P_e}}{\partial\bm{\tilde{\delta}}})_{coi}^{\star}}&=&\left[ \begin{array}{cc} 5.853 & 0.976\\ 3.524 & 5.289 \end{array}\right]
\end{eqnarray}
where the overline denotes the values after the contingency.

The actual $\overline{(\frac{\partial\bm{P_e}}{\partial\bm{\tilde{\delta}}})_{coi}}$ for the deterministic model without stochastic perturbations is the constant matrix as below:
\begin{equation}
\overline{(\frac{\partial\bm{P_e}}{\partial\bm{\tilde{\delta}}})_{coi}}=\left[ \begin{array}{cc}  5.943 &  0.949\\ 3.897 & 5.191\end{array}\right]
\end{equation}

By comparing the distances between $\overline{(\frac{\partial\bm{P_e}}{\partial\bm{\tilde{\delta}}})_{coi}}$ and the estimated matrixes, we have:
\begin{eqnarray}
\frac{\|\overline{(\frac{\partial\bm{P_e}}{\partial\bm{\tilde{\delta}}})_{coi}}-\overline{(\frac{\partial\bm{P_e}}{\partial\bm{\tilde{\delta}}})_{coi}^{\star}}\|_F}{\|\overline{(\frac{\partial\bm{P_e}}{\partial\bm{\tilde{\delta}}})_{coi}}\|_F}&=&4.48\%\\
\frac{\|\overline{(\frac{\partial\bm{P_e}}{\partial\bm{\tilde{\delta}}})_{coi}}-(\frac{\partial\bm{P_e}}{\partial\bm{\tilde{\delta}}})_{coi}\|_F}{\|\overline{(\frac{\partial\bm{P_e}}{\partial\bm{\tilde{\delta}}})_{coi}}\|_F}&=&27.07\%
\end{eqnarray}
which clearly demonstrate that the proposed hybrid method provides more accurate estimation for the Jacobian matrix after the contingency, because its performance is not affected by the change of network topology and parameter values.

From this example, some important insights can be obtained. The proposed method is able to provide accurate estimation for the system dynamic state Jacobian matrix %incorporates the stochastic dynamics induced by the variation of power injections and
by exploiting the statistical properties of the stochastic system. In addition, the performance of the proposed method outstands under imprecise knowledge or undetectable change of network topology and parameter values.

The state matrix $A$ that can be readily constructed from the Jacobian matrix $(\frac{\partial\bm{P_e}}{\partial\bm{\tilde{\delta}}})_{coi}$ provides uttermost important information on system conditions and dynamics that can be utilized in various ways. Firstly, the state matrix plays vital role in small signal stability analysis by estimating the oscillation frequency, mode shape and damping. In particular, the participation factor of the excited mode can be extracted, which measures the relative participation of the state variables, and thus facilitates locating oscillation source and designing countermeasures \cite{Kundur:book}. Secondly, the critical eigenvalue of the state matrix can be used as a measure of proximity to instability as discussed in extensive literature (e.g. \cite{Kundur:1992}-\cite{Cutsem:book}).
More importantly, the eigenvectors of the critical eigenvalue provide valuable information on the nature of the bifurcation, the response of the system and the control design \cite{Cutsem:book}.
%Apart from these, the Jacobian matrix can be used to locate the loadability surface with respect to the current operating point. In normal operation conditions, this will provide a security margin or protective control actions, whereas during emergency conditions, this will provide a direction for the corrective actions necessary to restore the operating point \cite{Cutsem:book}.
%The stability boundary can be either Hopf bifurcating points which lead to oscillations or saddle-node bifurcating points which result in system collapse.

\section{applications}\label{sectionapplications}
In this section, we describe two applications of the estimated Jacobian matrix and state matrix in online oscillation analysis and online stability monitoring and control respectively.
In addition, we show that the equation (\ref{rww}) can be used to approximate the damping of generators in model validation.

\subsection{Online Oscillation Analysis and Control}
We firstly show that the estimated state matrix by the proposed hybrid method can provide valuable information facilitating online power system oscillation analysis. Once the system state matrix is available, the oscillation frequency, mode shape and damping ratio can be immediately extracted. More importantly, the left eigenvector and the participation factor of the excited mode can also be calculated, which are not achievable from other purely measurement-based approaches yet play an important role in oscillation diagnosis.

\begin{figure}[!ht]
\centering
\begin{subfigure}[t]{0.52\linewidth}
\includegraphics[width=1.8in ,keepaspectratio=true,angle=0]{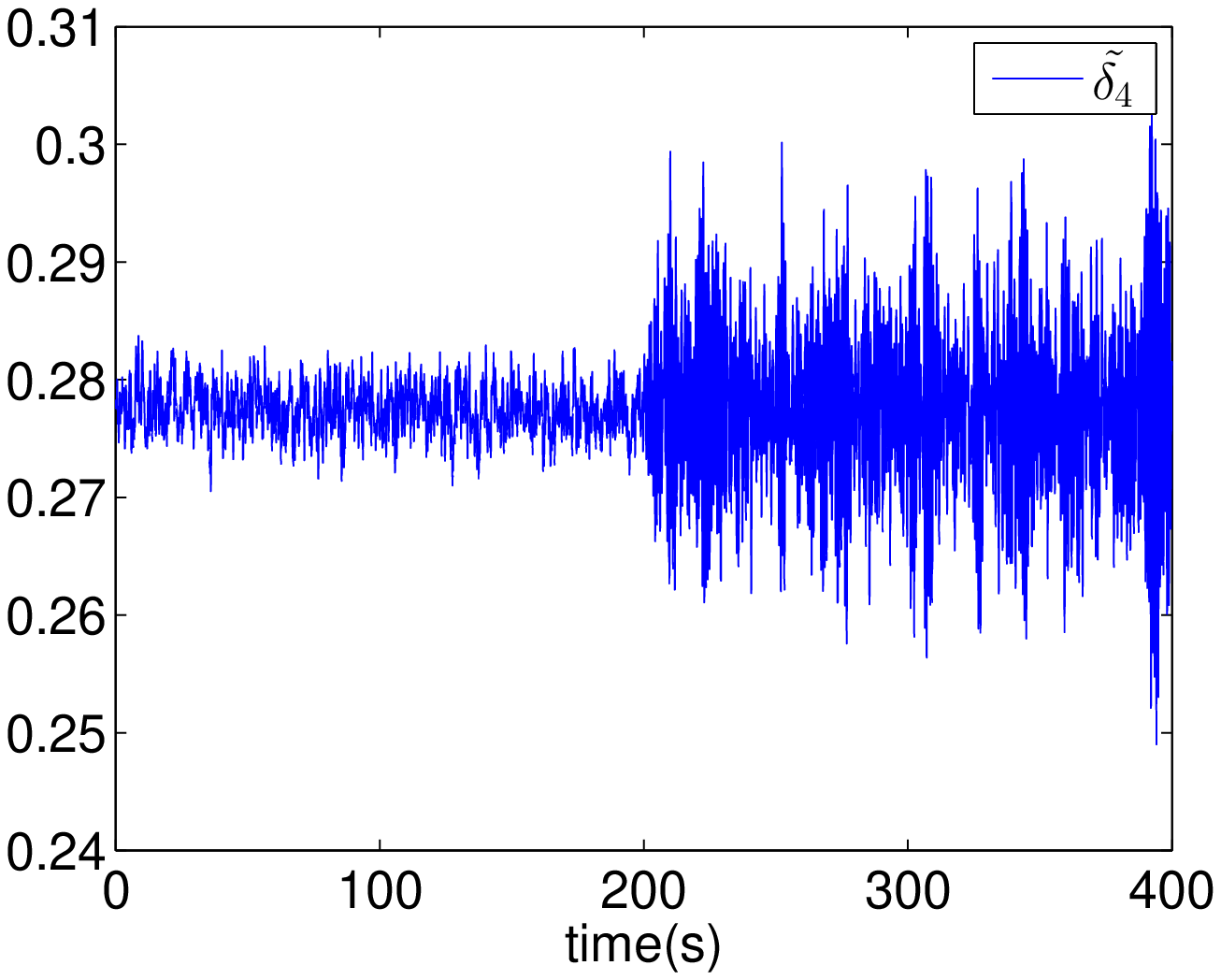}
\caption{Trajectory of $\tilde{\delta}_{4}$ on [0s,400s]}\label{d4-oscillation}
\end{subfigure}%
\begin{subfigure}[t]{0.48\linewidth}
\includegraphics[width=1.8in ,keepaspectratio=true,angle=0]{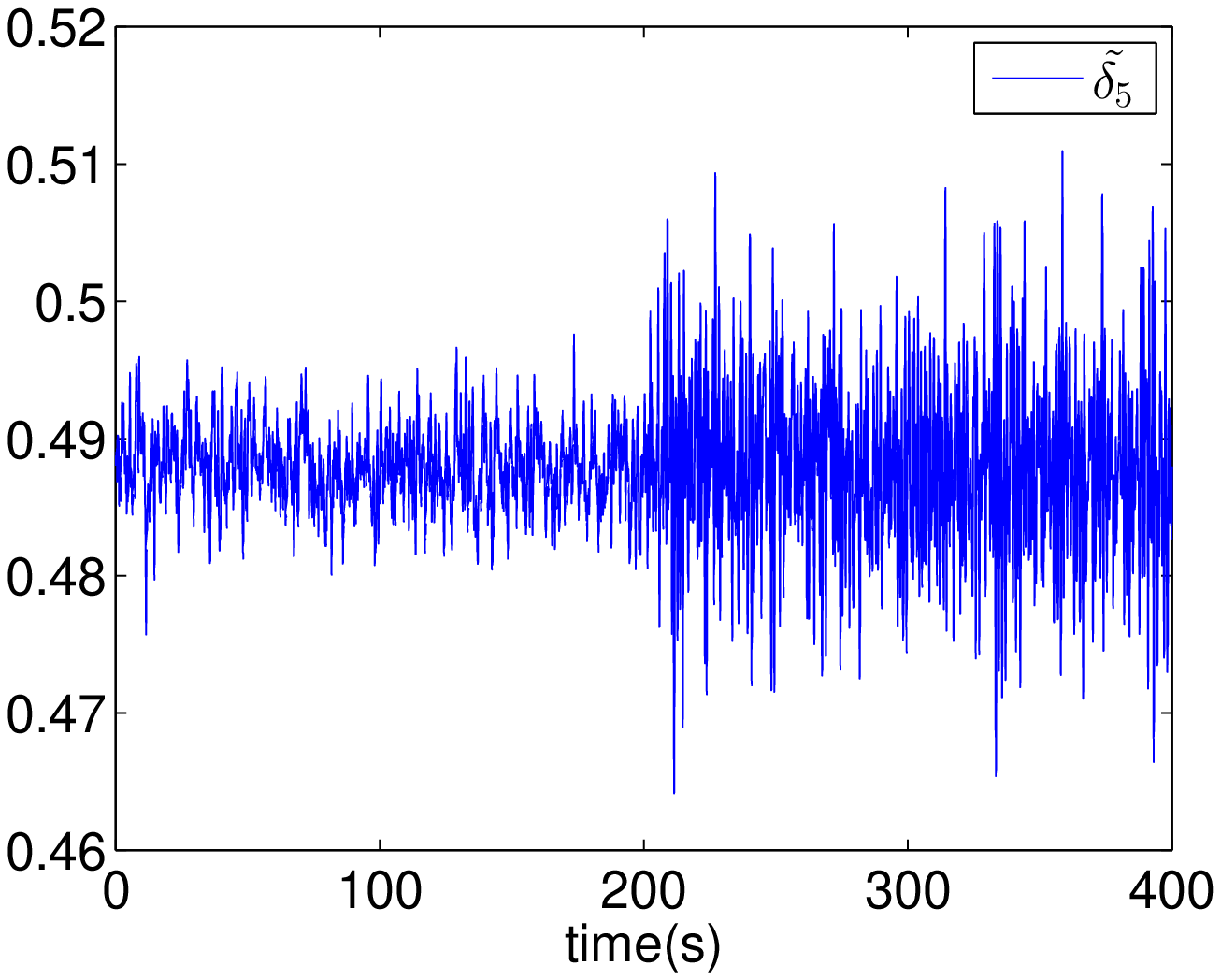}
\caption{Trajectory of $\tilde{\delta}_{5}$ on [0s,400s]}\label{d5-oscillation}
\end{subfigure}
\begin{subfigure}[t]{0.5\linewidth}
\includegraphics[width=1.8in ,keepaspectratio=true,angle=0]{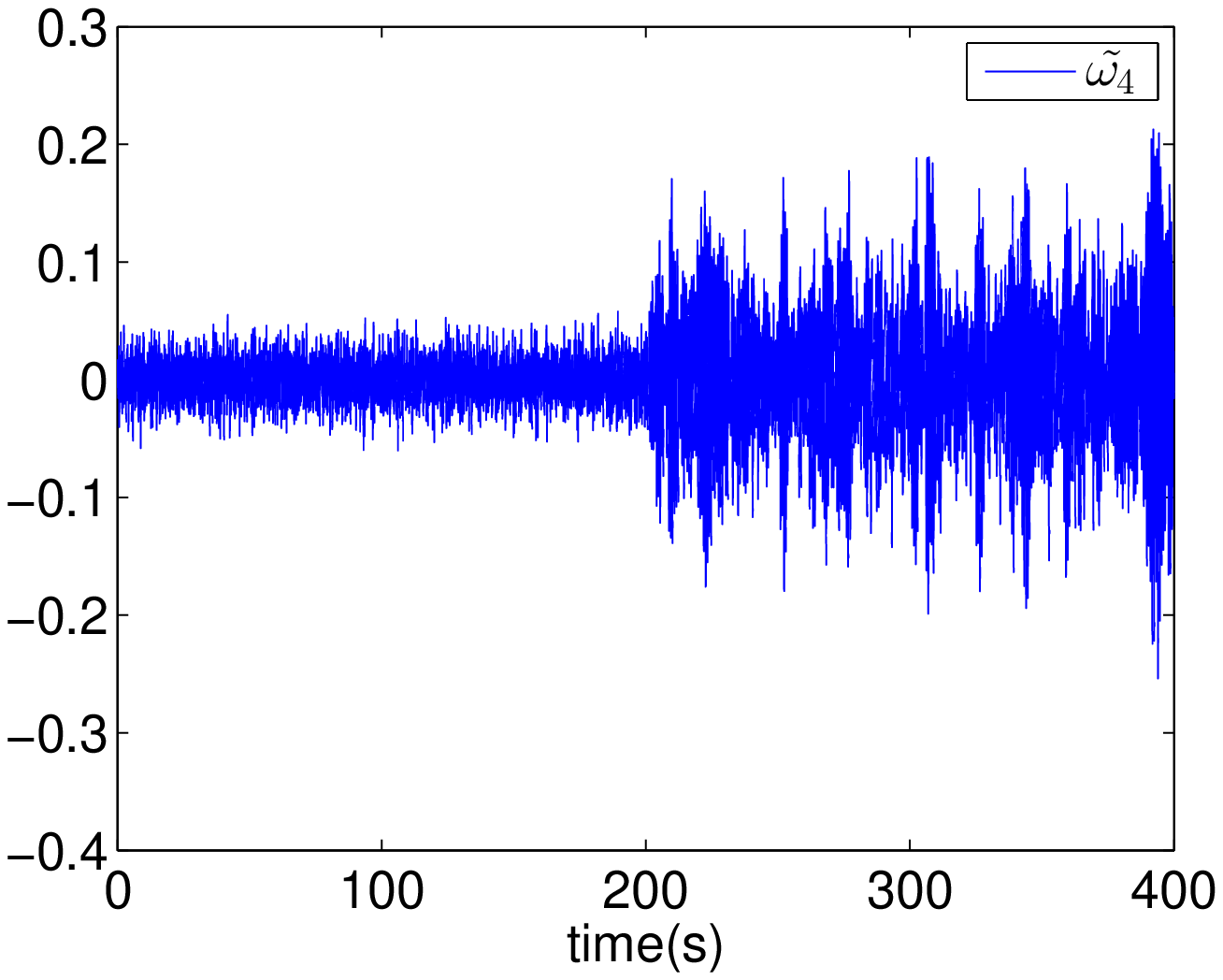}
\caption{Trajectory of $\tilde{\omega}_{4}$ on [0s,400s]}\label{w4-oscillation}
\end{subfigure}%
\begin{subfigure}[t]{0.5\linewidth}
\includegraphics[width=1.8in ,keepaspectratio=true,angle=0]{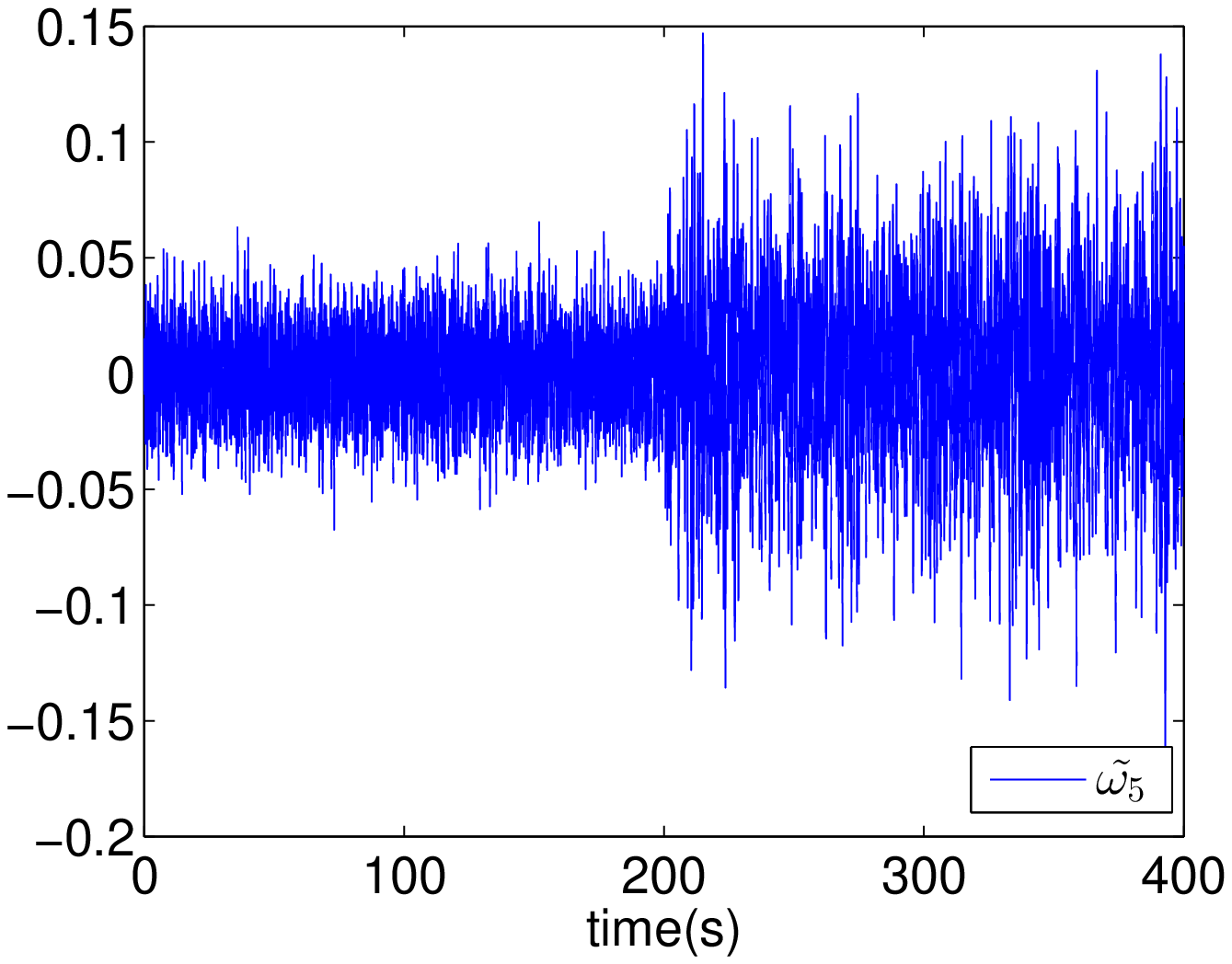}
\caption{Trajectory of $\tilde{\omega}_{5}$ on [0s,400s]}\label{w5-oscillation}
\end{subfigure}
\caption{Trajectories of $\tilde{\delta}_{4}$, $\tilde{\delta}_{5}$ and $\tilde{\omega}_{4}$, $\tilde{\omega}_{5}$ in the 39-bus system in COI reference.}\label{39-oscillation}
\end{figure}

A numerical example is given to illustrate the point. We consider the IEEE 39-bus test system that has 10 generators. The parameter values are available online: https://github.com/xiaozhew/Dynamic-State-Jacobian-Test-System. Assuming that the power injection at each generator experiences Gaussian noise with $\sigma_1=...=\sigma_{10}=0.01$. At 200s, a contingency occurs such that the system starts oscillating as shown in Fig. \ref{39-oscillation}. In order to investigate the oscillation and also diagnose the cause, we plan to utilize the Jacobian matrix and the state matrix estimated by the proposed hybrid method.

\begin{figure*}
\centering
\includegraphics[scale=0.56]{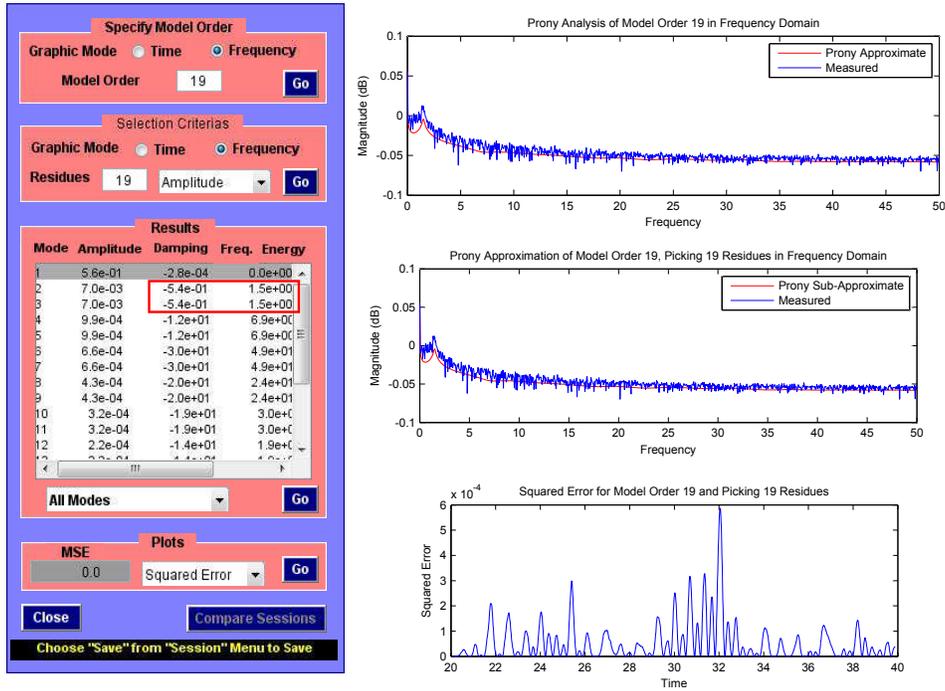}
\caption{Prony analysis for $\tilde{\delta}_{4}$ on [220s, 240s].} \label{prony}
\end{figure*}

By applying (\ref{proposemethod}), we are able to get $(\frac{\partial\bm{P_e}}{\partial\bm{\tilde{\delta}}})_{coi}$ after the contingency based on the measurements $\tilde{\bm{\delta}}$ and $\tilde{\bm{\omega}}$ from the PMUs, and furthermore construct the state matrix $A_{post}$ by (\ref{A}). The eigenvalues of $A_{post}$ are calculated and shown in Table \ref{39-oscillationeigenvalue}. It is seen that the complex conjugate pair ${-0.588 \pm 9.076i}$ have the lowest damping $-0.588$ with the oscillation frequency $\frac{9.076}{2\pi}=1.445$Hz, which is possibly the excited mode. %In order to verify this, the eigenvalue of the state matrix for the deterministic model is calculated and presented in Table \ref{39-oscillationeigenvalue-det}, showing that the excited mode is $-0.652 \pm 9.092i$.
In order to verify this, we carry out Prony analysis to the signal $\tilde{\delta}_4$ on [220s, 240s] by Prony Toolbox \cite{Pronytoolbox} and the results are presented in Fig. \ref{prony}. A 19th order model is used and 18 residues are identified by Prony analysis. It is marked by a red box in Fig. \ref{prony} that there is a frequency component at $1.5$ Hz whose damping is $-0.54$. The Prony analysis results are aligned with the eigenvalue analysis results.

\begin{table}[!ht]
\centering
\caption{Eigenvalues of $A_{post}$}\label{39-oscillationeigenvalue}
\begin{tabular}{|c|c|c|c|}
\hline
$\lambda_{1,2}$& $\mathbf{-0.588 \pm 9.076i}$&$\lambda_{3,4}$&$-1.070 \pm 5.769i$\\
\hline
$\lambda_{5,6}$&$-5.000\pm 8.548i$&$\lambda_{7,8}$&$ -5.000 \pm 8.762i$\\
\hline
$\lambda_{9}$& $-7.354$&$\lambda_{10}$&$-2.457$\\
\hline
$\lambda_{11,12}$&$-4.996 \pm 6.650i$&$\lambda_{13,14}$&$-4.999 \pm 6.370i$\\
\hline
$\lambda_{15,16}$&$-4.971 \pm 4.302i$&$\lambda_{17,18}$&$-4.971 \pm 4.860i$\\
\hline
\end{tabular}
\end{table}

In addition, the left eigenvector and thus the participation factor of the excited mode can be computed, which are not achievable by Prony analysis and other model-free methods.  %the eigenvalue analysis can provide eigenvectors, participation factor (i.e., eigenvalue sensitivity) that the Prony analysis can not.
%The participation factor indicates the relative participation of the respective states in the corresponding modes which can be utilized to locate oscillation source or problematic component \cite{Kundur:book}.
In this example, the participation factor for the mode $-0.588+9.076i$ is shown in Table \ref{39-oscillationPF}. It is observed that the maximum components are 0.436 at state $\tilde{\delta}_4$ and 0.437 at state $\tilde{\omega}_4$, implying that Generator 4 participates the most in this excited mode.

The actual cause for the oscillation is as follows. At 200s, some contingencies occur inside Generator 4 and 5 (e.g. faults within their control loops) such that the equivalent damping of Generator 4 decreases by 9 times and that of Generator 5 decreases by 4 times. The sudden decreasing of damping leads to the excitation of electromechanical oscillations, and both state variables of Generator 4 and 5 start oscillating. However, without detailed eigenvalue analysis, it can be hardly identified that Generator 4 takes the most responsibility in this excited oscillation mode, the fault inside which needs to be fixed in order to suppress the oscillation. %In fact, the excited mode in the deterministic model is $-0.652 + 9.092i$, whose participation factor has the maximum component 0.426 at both $\tilde{\delta}_4$ and $\tilde{\omega}_4$. We can see that the proposed method is able to capture the excited mode and meanwhile accurately identify the problematic component through participation factor. %which can not be obtained from other model-free approaches.
%However, without detailed eigenvalue analysis, it can be hardly identified that Generator 4 takes the most responsibility in this excited oscillation mode, the fault inside which needs to be fixed in order to suppress the oscillation.
\begin{table}[!ht]
\centering
\caption{Participation Factor of $-0.588+9.076i$}\label{39-oscillationPF}
\begin{tabular}{|c|c|c|c|c|}
\hline
$\tilde{\delta}_1$& $\tilde{\delta}_2$&$\tilde{\delta}_3$&$\tilde{\delta}_4$&$\tilde{\delta}_5$\\
\hline
3.22e-04&1.48e-04&1.04e-03&\textbf{4.36e-01}&5.96e-02\\
\hline
$\tilde{\delta}_6$& $\tilde{\delta}_7$&$\tilde{\delta}_8$&$\tilde{\delta}_9$&\\
\hline
{1.84e-03}&{7.48e-04}&{2.39e-04}&{5.47e-04}&\\
\hline
$\tilde{\omega}_1$&$\tilde{\omega}_2$& $\tilde{\omega}_3$&$\tilde{\omega}_4$&$\tilde{\omega}_5$\\
\hline
{2.24e-04}&{1.03e-04}&{7.21e-04}&\textbf{{4.37e-01}}&{5.90e-02}\\
\hline
$\tilde{\omega}_6$&$\tilde{\omega}_7$& $\tilde{\omega}_8$&$\tilde{\omega}_9$&\\
\hline
{1.28e-03}&{5.20e-04}&{1.66e-04}&{3.81e-04}&\\
\hline
\end{tabular}
\end{table}

This example shows that the proposed hybrid method enables eigenvalue analysis for power system oscillations without detailed knowledge on network parameters. The analysis can provide comprehensive information %including oscillation modes, mode shape, damping, participation factor, etc.
with respect to the oscillation in near real-time, making the online oscillation analysis accurate and effective. In particular, the proposed method can provide participation factor which is unobtainable from other model-free approaches yet is of great significance in oscillation diagnosis and control.

\subsection{Online Stability Monitoring and Control}
Another application of the estimated state matrix is to use its critical eigenvalue as a measure of proximity to instability. More importantly, the eigenvectors of the critical eigenvalue carries significant information on the behavior of the systems and the emergency control actions. A numerical example is given to illustrate this application.% of the estimated Jacobian in stability monitoring and control.

\begin{figure}[!ht]
\centering
\begin{subfigure}[t]{0.52\linewidth}
\includegraphics[width=1.8in ,keepaspectratio=true,angle=0]{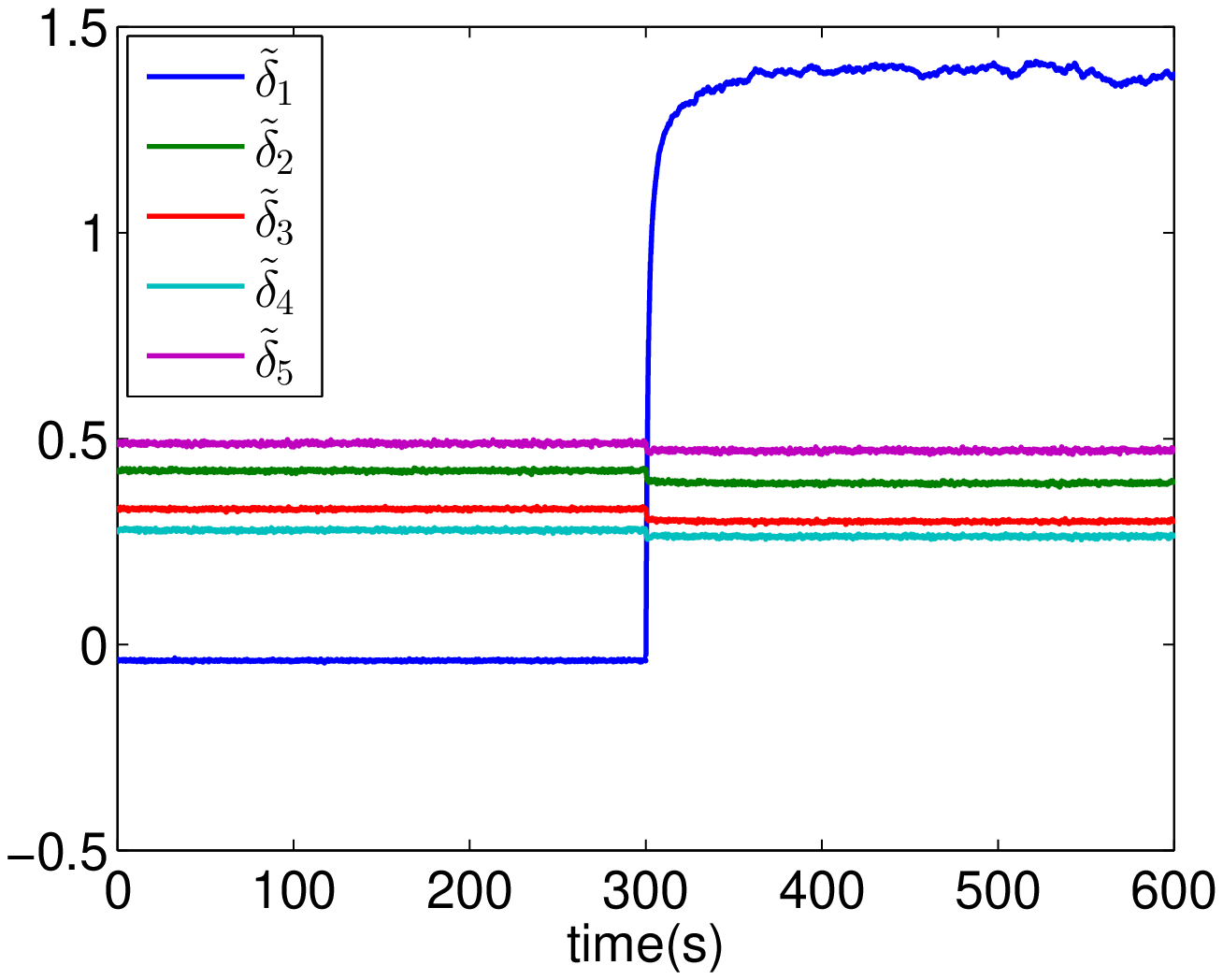}
\caption{Trajectories of $\tilde{\delta}_{1}$-$\tilde{\delta}_{5}$ on [0s,600s]}\label{d1-5-39}
\end{subfigure}%
\begin{subfigure}[t]{0.5\linewidth}
\includegraphics[width=1.8in ,keepaspectratio=true,angle=0]{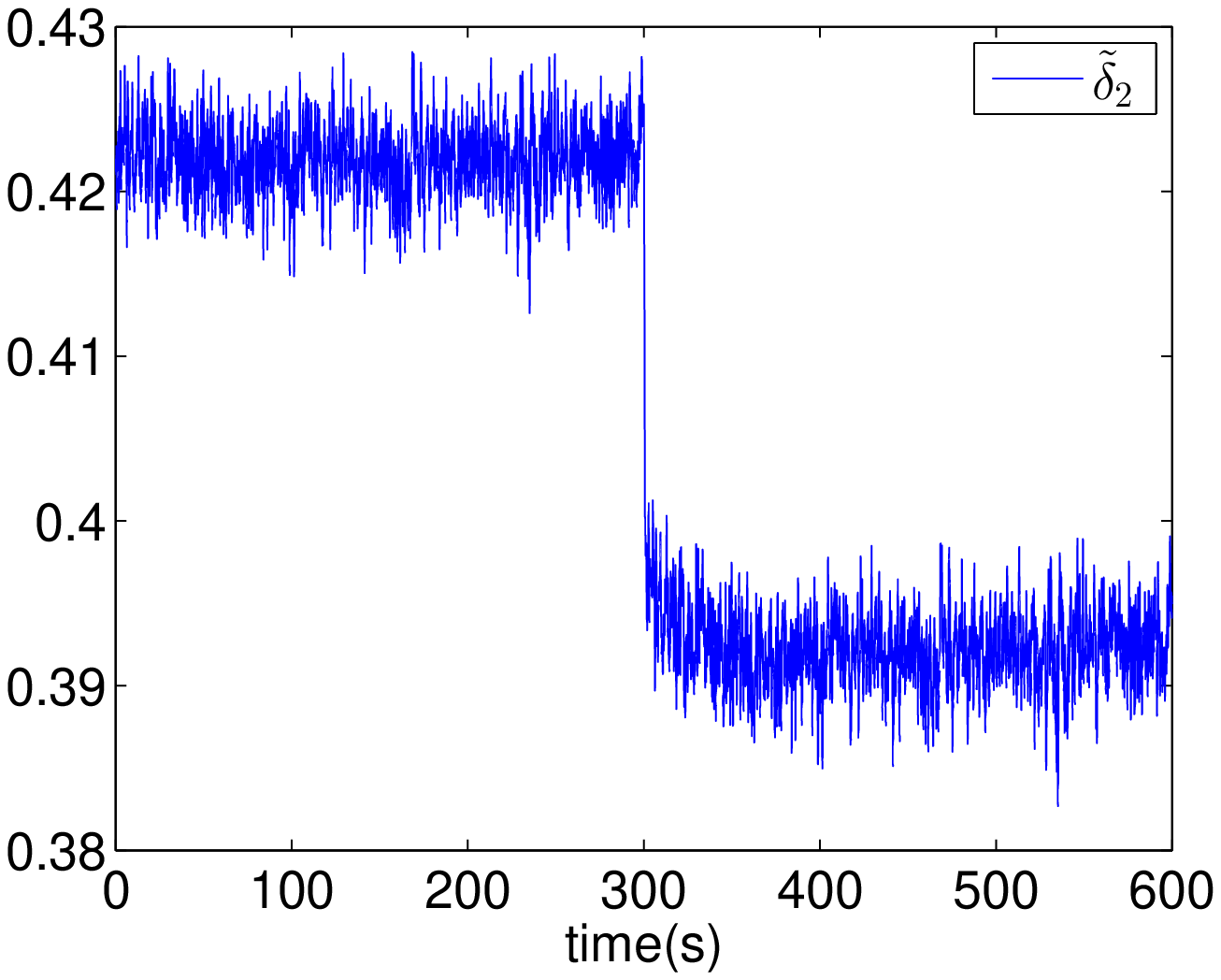}
\caption{Trajectory of $\tilde{\delta}_{2}$ on [0s,600s]}\label{d2-39-stb}
\end{subfigure}
\begin{subfigure}[t]{0.48\linewidth}
\includegraphics[width=1.8in ,keepaspectratio=true,angle=0]{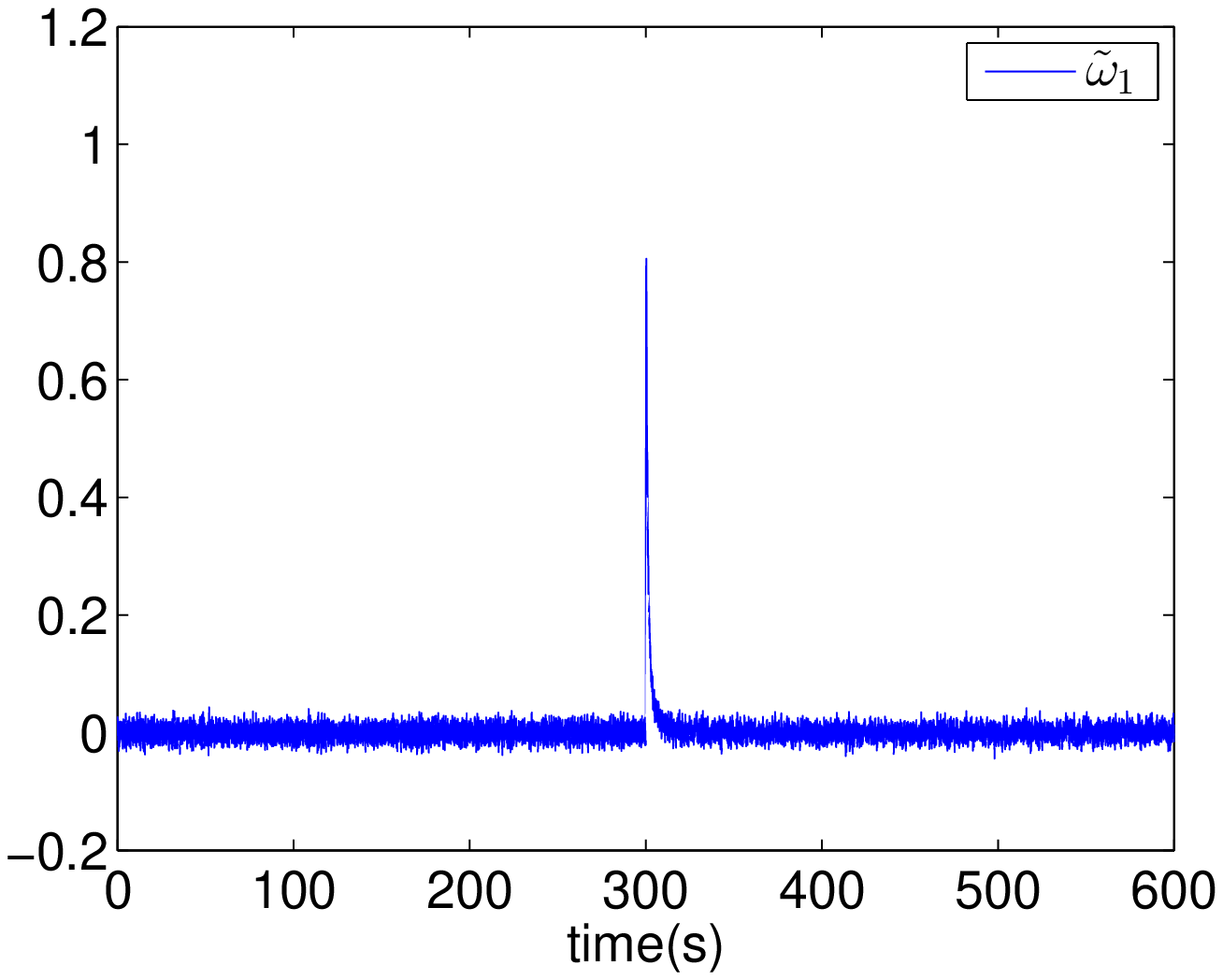}
\caption{Trajectory of $\tilde{\omega}_{1}$ on [0s,600s]}\label{w1-39-stb}
\end{subfigure}%
\begin{subfigure}[t]{0.5\linewidth}
\includegraphics[width=1.8in ,keepaspectratio=true,angle=0]{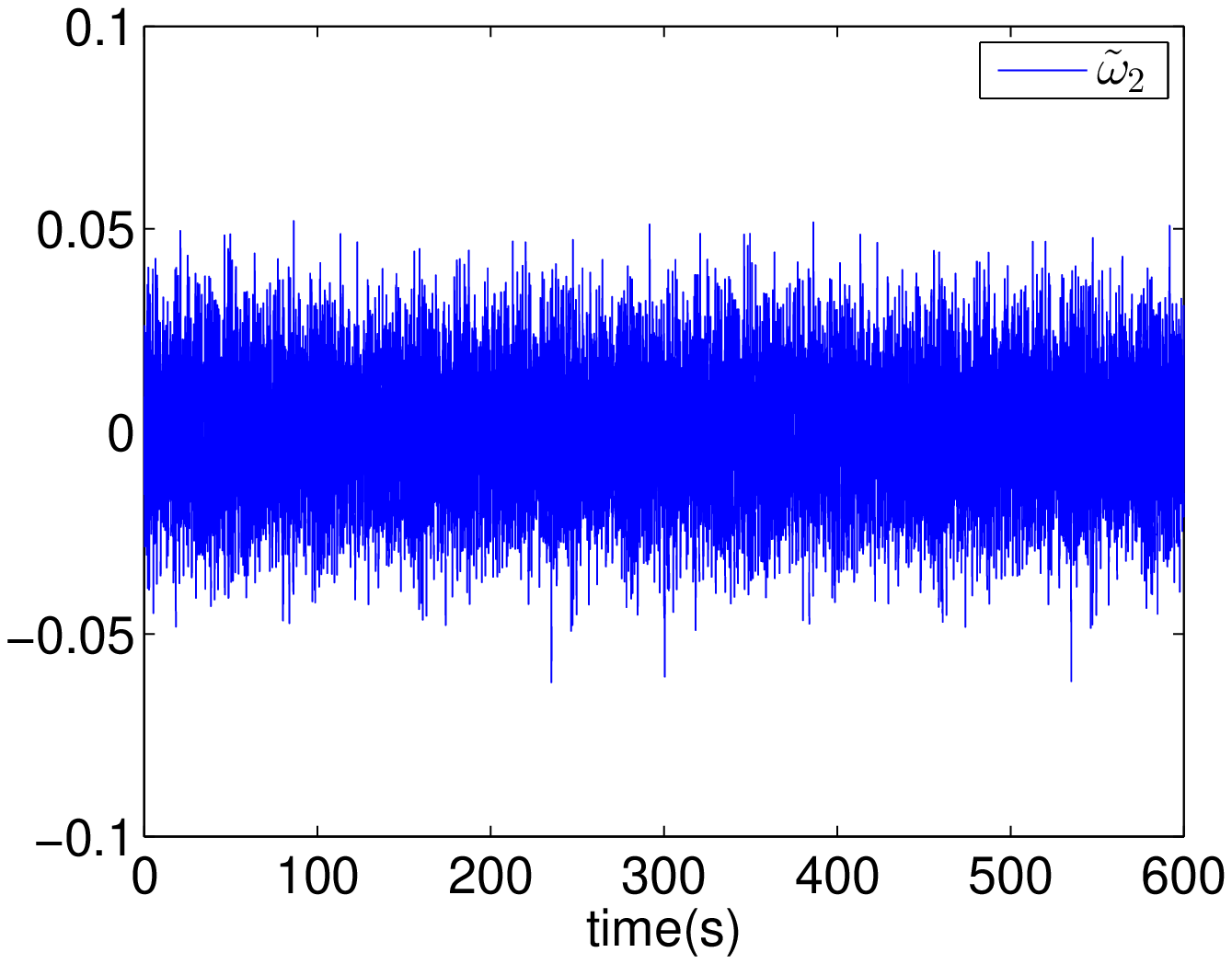}
\caption{Trajectory of $\tilde{\omega}_{2}$ on [0s,600s]}\label{w2-39-stb}
\end{subfigure}
\caption{The trajectories of some state variables in the 39-bus system in COI reference before and after the contingency.}\label{39-bus-stability}
\end{figure}

We also consider the IEEE 39-bus 10-generator system, the parameter values of which are the same as the last example. We assume that the variation of power injection at each machine is Gaussian with $\sigma_1=...=\sigma_{10}=0.01$.  At 300.01s, a contingency occurs inside Generator 1 such that its transient reactance $x_d'$ changes from $0.031$ p.u. to $0.4311$ p.u., as a result, the system has been pushed considerably close to its stability boundary. However, one can hardly  assess the degree of stability of the system directly from the trajectories of the state variables shown in Fig. \ref{39-bus-stability}.
%The trajectories of rotor angles for Generator 1-5 before and after the contingency are shown in Fig. \ref{d1-5-39}, specifically the zoomed-in trajectory of $\tilde{\delta}_{2}$ is shown in Fig \ref{d2-39-stb}. Fig. \ref{w1-39-stb}-\ref{w2-39-stb} present the trajectories of $\tilde{\omega}_{1}$ and $\tilde{\omega}_{2}$ respectively.
%One can hardly assess the degree of stability of the system directly from the trajectories shown in Fig. \ref{145-bus}, the critical eigenvalue of the dynamic state Jacobian matrix, however, carries such important information and is able to measure the degree of stability.

\begin{figure}[!ht]
\centering
\includegraphics[width=2.5in ,keepaspectratio=true,angle=0]{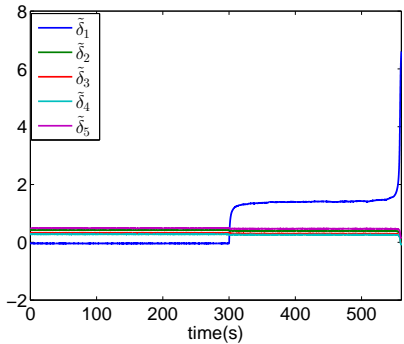}\caption{Trajectories of $\tilde{\delta}_{1}$-$\tilde{\delta}_{5}$ when the system loses the stability.}\label{d1-5-afterSNB}
\end{figure}

By applying the proposed hybrid method, we are able to construct the state matrix $A_{post}$ after the contingency from the PMU measurements $\bm{\tilde{\delta}}$ and $\bm{\tilde{\omega}}$. In particular, the critical eigenvalue of$A_{post}$ is -0.026, indicating that the system is fairly close to the saddle node bifurcation point, i.e., its stability boundary, due to the contingency. To verify this, we aggravate the contingency slightly by increasing $x_d'$ of Generator 1 from $0.4311$ p.u. to $0.4312$ p.u., and the trajectories of  $\tilde{\delta}_{1}$-$\tilde{\delta}_{5}$ are shown in Fig. \ref{d1-5-afterSNB} which clearly demonstrates that the system loses its stability.

We further compute the right eigenvector corresponding to the critical eigenvalue -0.026 %as the left and right eigenvectors carry useful information describing the system dynamics near the stability boundary \cite{Cutsem:book}. In particular, the right eigenvector corresponding to the critical eigenvalue indicates
that identifies %which generator angles will increase due to the bifurcation, and consequently
which machines are most likely to lose synchronism due to the bifurcation \cite{Cutsem:book}\cite{Vournas:1998}. The normalized right eigenvector of the critical eigenvalue is shown in Table \ref{39-stbRE}, implying that Generator 1 is going to lose synchronism, which exactly matches the simulation result in Fig. {\ref{d1-5-afterSNB}}.
%further validate the capability of the proposed hybrid method in stability monitoring and control.
%\begin{table}[!ht]
%\centering
%\caption{The Normalized Right Eigenvector of -0.025}\label{39-stbRE}
%\begin{tabular}{|c|c|c|c|c|}
%\hline
%$\delta_1$& $\delta_2$&$\delta_3$&$\delta_4$&$\delta_5$\\
%\hline
%-9.99e-01&2.31e-02&1.72e-02&7.75e-03&6.18e-03\\
%\hline
%$\delta_6$& $\delta_7$&$\delta_8$&$\delta_9$&\\
%\hline
%7.79e-03&{9.81e-03}&{4.58e-03}&{-1.85e-03}&\\
%\hline
%$\omega_1$&$\omega_2$& $\omega_3$&$\omega_4$&$\omega_5$\\
%\hline
%{2.56e-02}&{-5.94e-04}&{-4.41e-04}&-1.99e-04& -1.59e-04\\
%\hline
%$\omega_6$&$\omega_7$& $\omega_8$&$\omega_9$&\\
%\hline
%{ -2.00e-04}&{ -2.51e-04}&{-1.17e-04}&{4.73e-05}&\\
%\hline
%\end{tabular}
%\end{table}

\begin{table}[!ht]
\centering
\caption{The Normalized Right Eigenvector of -0.026}\label{39-stbRE}
\begin{tabular}{|c|c|c|c|c|}
\hline
$\tilde{\delta}_1$& $\tilde{\delta}_2$&$\tilde{\delta}_3$&$\tilde{\delta}_4$&$\tilde{\delta}_5$\\
\hline
\textbf{0.9991}&-0.0232&-0.0172&-0.0077&-0.0059\\
\hline
$\tilde{\delta}_6$& $\tilde{\delta}_7$&$\tilde{\delta}_8$&$\tilde{\delta}_9$&\\
\hline
-0.0076&-0.0098&-0.0046&0.0019&\\
\hline
$\tilde{\omega}_1$&$\tilde{\omega}_2$& $\tilde{\omega}_3$&$\tilde{\omega}_4$&$\tilde{\omega}_5$\\
\hline
-0.0256&0.0006&0.0004&0.0002&0.0002\\
\hline
$\tilde{\omega}_6$&$\tilde{\omega}_7$& $\tilde{\omega}_8$&$\tilde{\omega}_9$&\\
\hline
0.0002&0.0003&0.0001&-0.00004&\\
\hline
\end{tabular}
\end{table}

In addition to the right eigenvector, the left eigenvector of the critical eigenvalue provides essential information on the emergency control design. In particular, the left eigenvector in the state space is related to the normal vector to the bifurcation surface in parameter space \cite{Cutsem:book}\cite{Dobson:1992}\cite{Dobson:1993}. For instance, we want to re-dispatch the generators in order to push the system back to the secure region once we realize that the system has been pushed extremely close to its stability boundary.

\begin{figure}[!ht]
\centering
\includegraphics[width=1.8in ,keepaspectratio=true,angle=0]{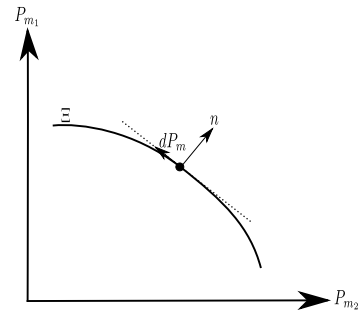}\caption{An illustration of the normal vector $\bm{n}$ in the parameter space.}\label{normal vector}
\end{figure}

The power system model (\ref{swing-1})-(\ref{swing-2}) can be represented as:
\begin{equation}
\bm{\dot{x}}=\bm{h}(\bm{x},\bm{P_m})\label{SNBparameter}
\end{equation}
where $\bm{x}=[\bm{\delta},\bm{\omega}]^T$, and regard $\bm{P_m}$ as parameters. The saddle node bifurcation surface $\Xi$ in parameter space $\bm{P_m}\in\mathbb{R}^n$ is the set of $\bm{P_m}$ that yields a saddle node bifurcation of (\ref{SNBparameter}). Near the bifurcation surface, we have:
\begin{equation}
\frac{\partial{\bm{h}}}{\partial{\bm{x}}}\bm{dx}+\frac{\partial{\bm{h}}}{\partial{\bm{P_m}}}\bm{dP_m}=0
\end{equation}
Premultiplying by $\bm{w}^T$, where $\bm{w}$ is the left eigenvector corresponding to the zero eigenvalue of $\frac{\partial{\bm{h}}}{\partial{\bm{x}}}$, we obtain:
\begin{equation}
\bm{w}^T\frac{\partial{\bm{h}}}{\partial{\bm{x}}}\bm{dx}+\bm{w}^T\frac{\partial{\bm{h}}}{\partial{\bm{P_m}}}\bm{dP_m}=\bm{w}^T\frac{\partial{\bm{h}}}{\partial{\bm{P_m}}}\bm{dP_m}=0
\end{equation}
Specifically, $\bm{n}=(\frac{\partial{\bm{h}}}{\partial{\bm{P_m}}})^T\bm{w}$ such that $\bm{n}^T\bm{dP_m}=0$. $\bm{n}$ is orthogonal to any small $\bm{dP_{m}}$ lying on $\Xi$, and is thus the normal vector of $\Xi$ at the point considered as illustrated in Fig. \ref{normal vector}.

\begin{figure}[!ht]
\centering
\begin{subfigure}[t]{0.52\linewidth}
\includegraphics[width=1.8in ,keepaspectratio=true,angle=0]{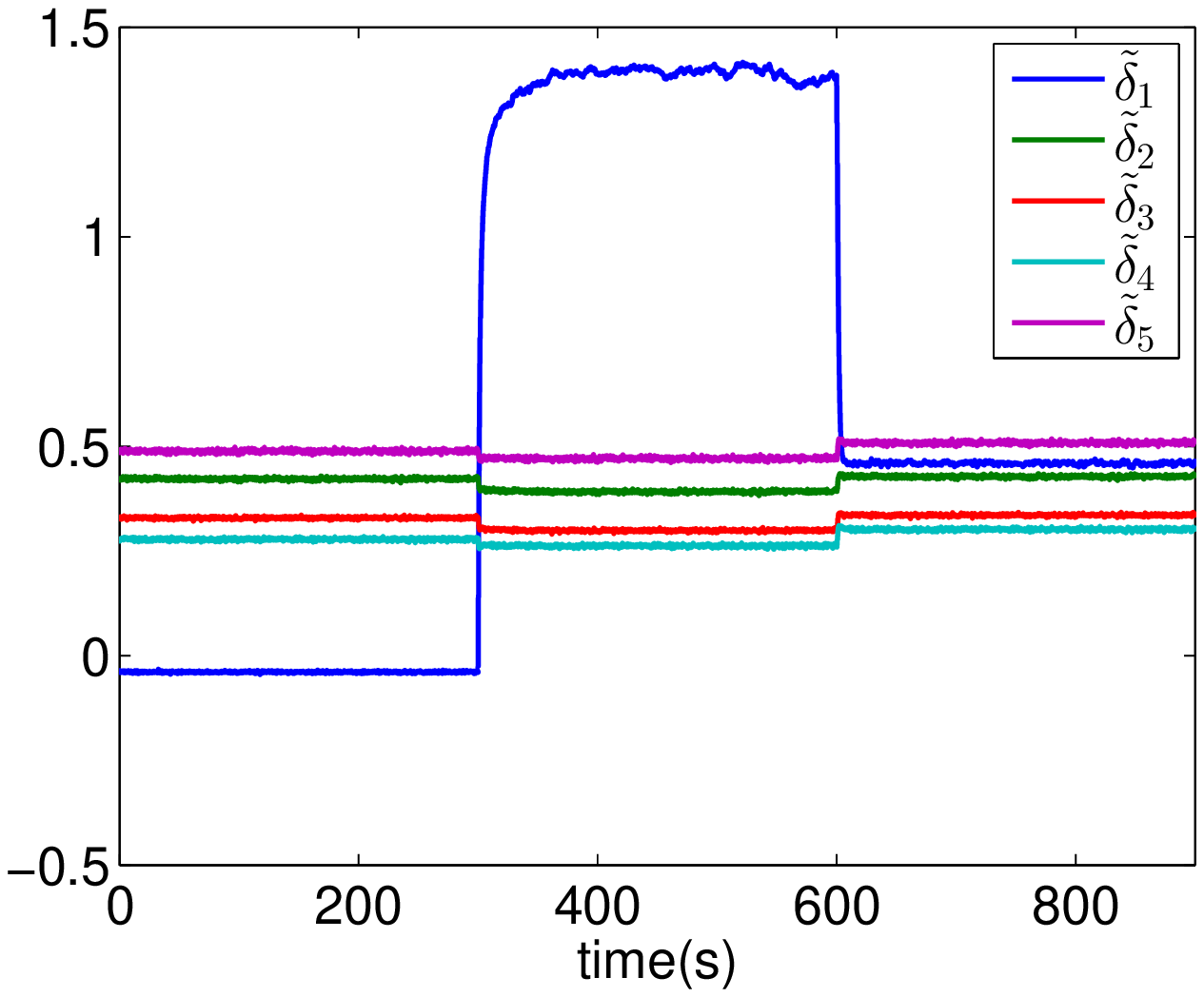}
\caption{Trajectories of $\tilde{\delta}_{1}$-$\tilde{\delta}_{5}$ on [0s,900s]}\label{d1-5-39-control}
\end{subfigure}%
\begin{subfigure}[t]{0.5\linewidth}
\includegraphics[width=1.8in ,keepaspectratio=true,angle=0]{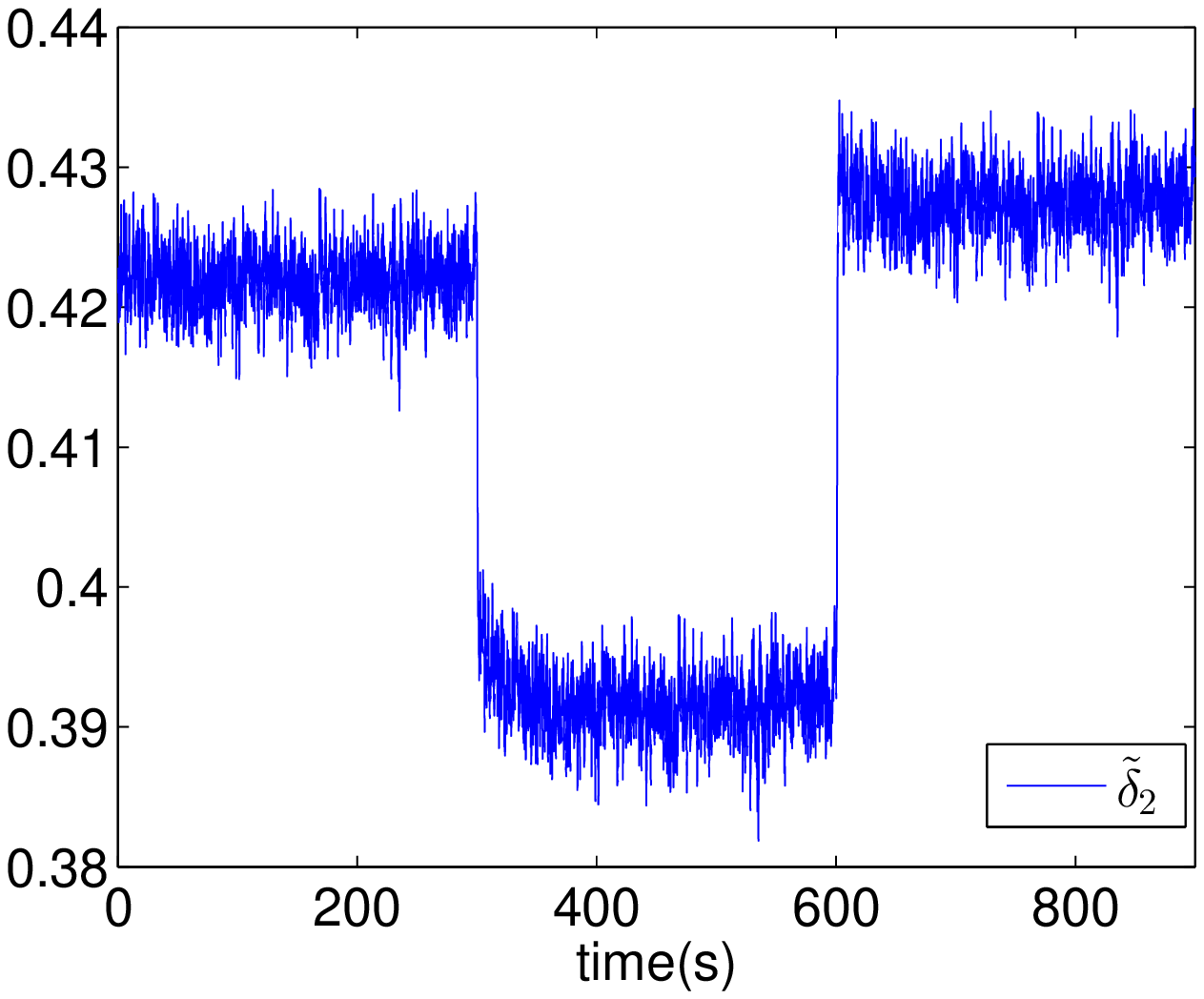}
\caption{Trajectory of $\tilde{\delta}_{2}$ on [0s,900s]}\label{d2-39-stb-control}
\end{subfigure}
\caption{The trajectories of some state variables in the 39-bus system before and after the contingency as well as after the generation re-dispatch.}\label{39-bus-control}
\end{figure}

In this case, the unit normal vector at the bifurcation point on the surface is shown in Table \ref{39-stbNV}. In order to push the system back to its stability margin, let $\bm{P_m}$ move towards the direction $-{\bm{n}}$ for one unit, and make the Generator 9 pick up the resulting decreasing power 0.9766 p.u., as it is the least sensitive component in this unstable mode. Suppose this emergency generation re-dispatch takes place at 600s, then the system trajectories are shown in Fig. \ref{39-bus-control}, from which it is seen that the operation point moves towards the safe region. Additionally, the critical eigenvalue of the state matrix is also computed to further judge the effectiveness of the control. It is shown that the critical eigenvalue changes from -0.026 to -1.131 owing to the control action, demonstrating that the generation re-dispatch has effectively moved the system back to its stability region.

\begin{table}[!ht]
\centering
\caption{The Unit Normal Vector of -0.026}\label{39-stbNV}
\begin{tabular}{|c|c|c|c|c|}
\hline
$P_{m_1}$& $P_{m_2}$&$P_{m_3}$&$P_{m_4}$&$P_{m_5}$\\
\hline
\textbf{0.9995}&-0.0080&-0.0087&-0.0063& 0.0104\\
\hline
$P_{m_6}$& $P_{m_7}$&$P_{m_8}$&$P_{m_9}$&\\
\hline
0.0138&-0.0200& -0.0060&0.0018&\\
\hline
\end{tabular}
\end{table}

%According to the model-based method, the critical eigenvalue of the state matrix $A_{post}$ for the deterministic model is -0.007, the right eigenvector of which has the maximum component 0.9994 at $\tilde{\delta}_1$, and the normal unit vector of which is presented in Table
%\begin{table}[!ht]
%\centering
%\caption{The Unit Normal Vector of -0.007 in the Deterministic Model}\label{39-stbNV-model}
%\begin{tabular}{|c|c|c|c|c|}
%\hline
%$P_{m_1}$& $P_{m_2}$&$P_{m_3}$&$P_{m_4}$&$P_{m_5}$\\
%\hline
%\textbf{0.9991}&-0.0161&-0.0153& -0.0128&-0.0197\\
%\hline
%$P_{m_6}$& $P_{m_7}$&$P_{m_8}$&$P_{m_9}$&\\
%\hline
% -0.0162& -0.0126&  0.0008& -0.0185&\\
%\hline
%\end{tabular}
%\end{table}

This example illustrates one of the important applications of the state matrix that is achievable by the proposed method. Firstly, the critical eigenvalue of the estimated $A$ is a good measure of proximity to instability for a system, the accuracy of which has been well demonstrated by the simulation results. More than a simple stability indicator, the eigenvectors of the critical eigenvalue are able to provide comprehensive and valuable information on the response of the system after the bifurcation and the means of emergency control. Simulation results show that the emergency generation re-dispatch designed based on the left eigenvector can effectively move the system back to its stability region. As the hybrid model and measurement-based method can acquire the state matrix in near real-time, we believe that it can play an important role and provide significant information for online stability monitoring and control.

\subsection{Model Validation}
Since models are the foundation of most power system studies, power system model validation is an essential procedure for maintaining system security and reliability, which should be done periodically \cite{Bialek:2014}\cite{Pourbeik:2010}. Dynamic equivalencing, that is, obtaining a reduced-order power system model to capture the relevant dynamics, has been an active research area to reduce the computational effort of dynamic security assessment. In particular, the classical aggregation method has been widely implemented in practice. By this method, the equivalenced generator is formed using a classical representation \cite{Vittal:2013}. In this part, we will show that the derived relation (\ref{rww}) can help estimate the equivalent damping $D$ for generators.

\begin{figure}[!ht]
\centering
\begin{subfigure}[t]{0.5\linewidth}
\includegraphics[width=1.8in ,keepaspectratio=true,angle=0]{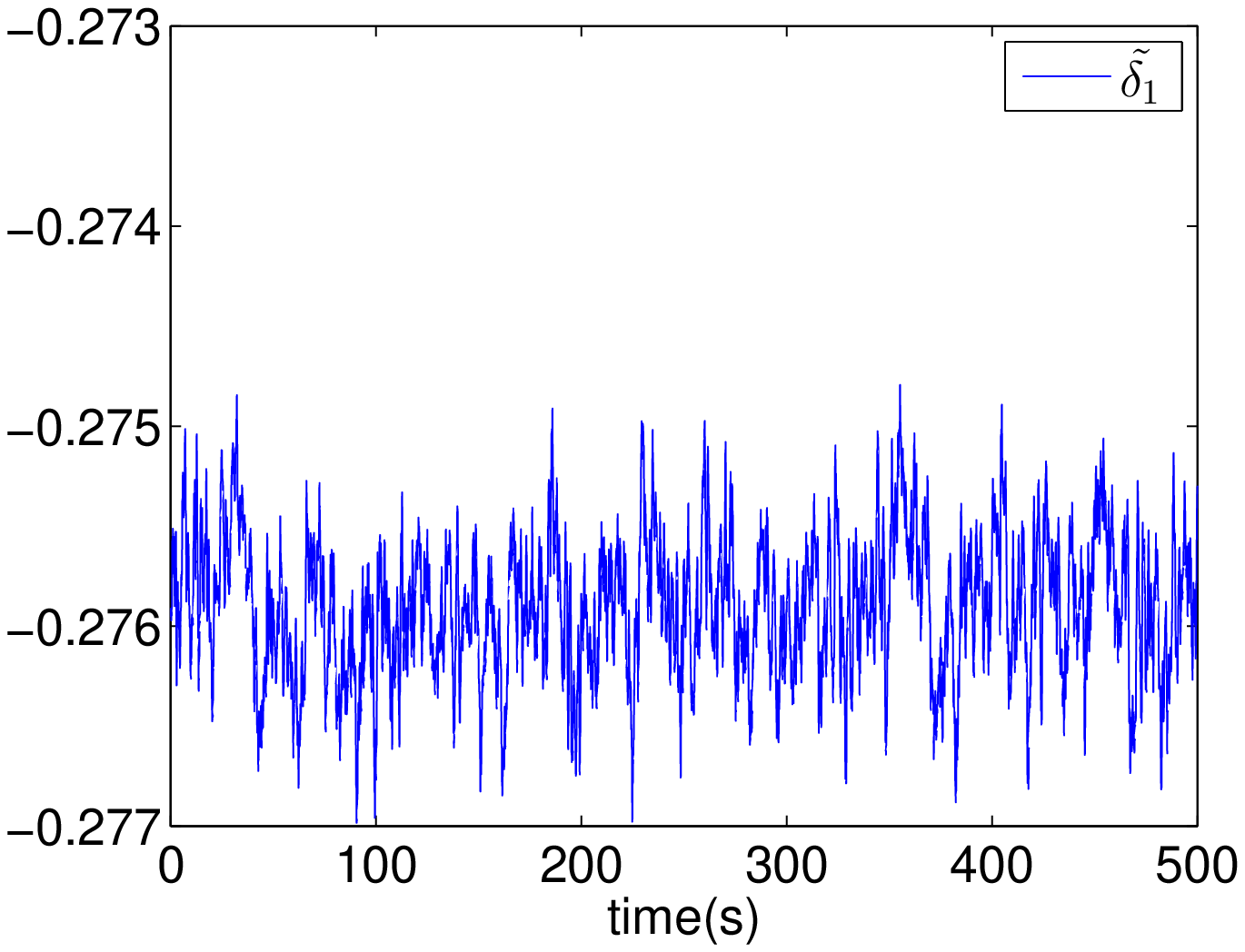}
\caption{Trajectory of $\tilde{\delta}_{1}$ on [0s,500s]}\label{d1-39}
\end{subfigure}%
\begin{subfigure}[t]{0.5\linewidth}
\includegraphics[width=1.8in ,keepaspectratio=true,angle=0]{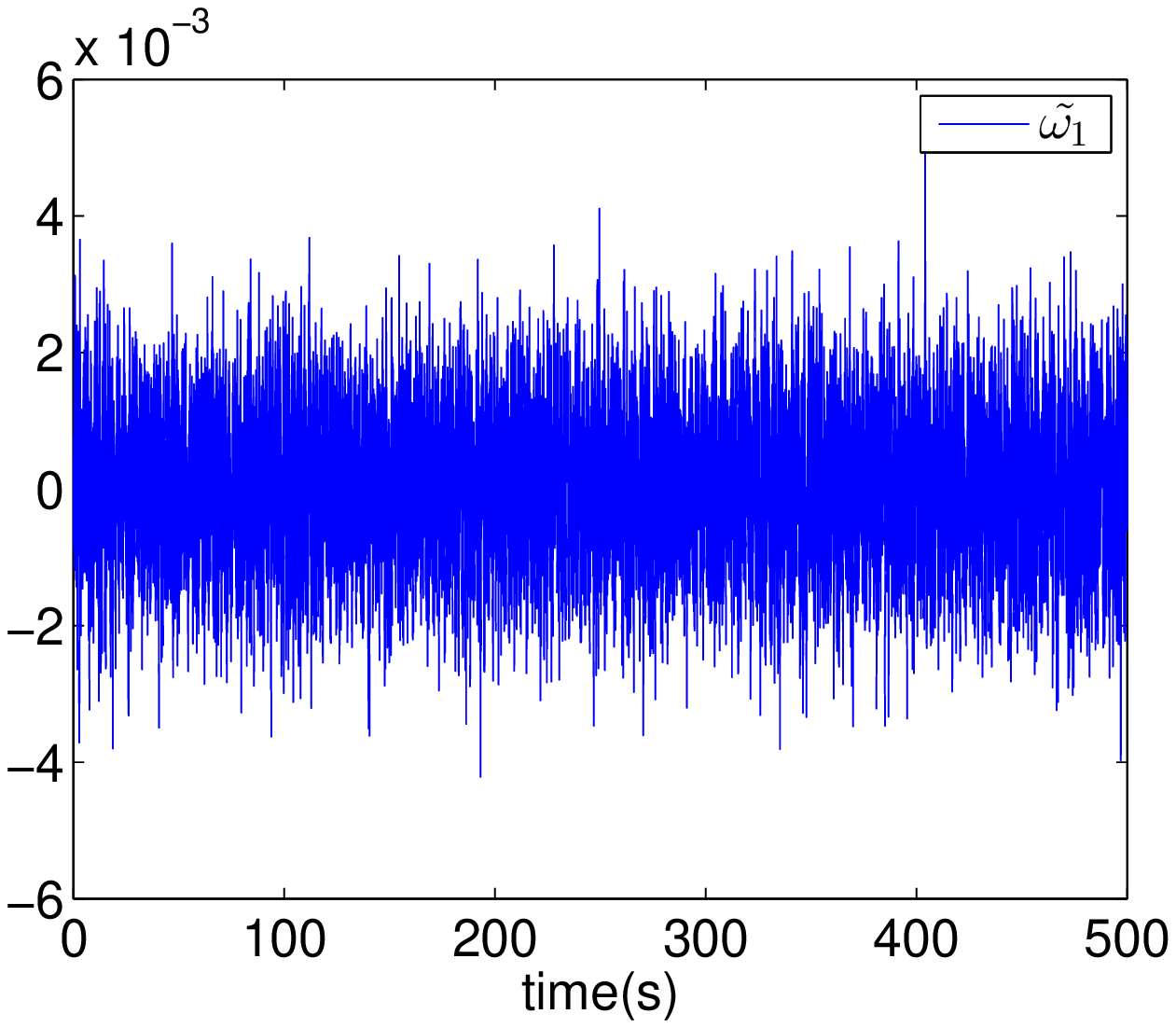}
\caption{Trajectory of $\tilde{\omega}_{1}$ on [250s,400s]}\label{w1-39}
\end{subfigure}
\caption{The trajectories of some state variables in the 39-bus system in COI reference.}\label{39-modelvalidation}
\end{figure}

We consider the IEEE 39-bus test system, the parameter values of which are available online: https://github.com/xiaozhew/Dynamic-State-Jacobian-Test-System. In order to approximate $D$, we add a Gaussian noise to the power injection at each generator with $\sigma_1=...=\sigma_{10}=0.01$. Trajectories of some state variables on the time scale [0s 500s] are shown in Fig. \ref{39-modelvalidation}.

Suppose we know $M$, and is able to calculate $C_{\bm{\omega}{\bm{\omega}}}$ from the PMU measurements over the 500s, then by the relation (\ref{rww}), we have:
\begin{equation}
D=\frac{1}{2}\Sigma^2M^{-1}C_{\bm{\omega}{\bm{\omega}}}^{-1} \label{D}
\end{equation}

The actual damping and the estimated damping by the proposed method is shown in Table \ref{39-dampingtable}. It can be found that the proposed method can estimate the damping for generators in reasonably good accuracy.
\begin{table}[!ht]
\centering
\caption{The comparison between the actual and estimated damping}\label{39-dampingtable}
\begin{tabular}{|c|c|c|}
\hline
actual damping&estimated damping&error\\
\hline
22.28&21.52&3.41\%\\
\hline
16.07&15.18&5.54\%\\
\hline
18.99&17.86&5.95\%\\
\hline
15.17&14.35&5.34\%\\
\hline
13.79&13.49&2.18\%\\
\hline
18.46&17.52&5.09\%\\
\hline
14.01&13.10&6.50\%\\
\hline
12.89&12.04&6.59\%\\
\hline
18.30&17.54&4.15\%\\
\hline
26.53&25.06&5.54\%\\
\hline
\end{tabular}
\end{table}

\section{Conclusions and Perspectives}\label{sectionconclusion}

In this paper, we have proposed a hybrid measurement and model-based method for estimating dynamic state Jacobian matrix in near real-time. The proposed hybrid method works as a grey box bridging the measurement and the model, and is able to provide fairly accurate estimation without being affected by the variation of network topology. The estimated Jacobian matrix has various applications including online oscillation analysis, stability monitoring and emergency control. A side-product of the method also provides an alternative way to estimate damping for the generators in model validation. In the future, we plan to explore other applications of the estimated Jacobian matrix in power system operation such as congestion relief, economic dispatch and preventive control design. Further investigations of the method on higher-order generator models and detailed load models are expected. %An extension of the method for higher order model is also expected.

\appendices
\section{validation on higher order model}\label{higherordermodel}
In this paper, we focus on ambient oscillations around stable steady state, where classical generator model can reasonably represent the system dynamics. Nevertheless, it can be shown that applying higher-order generator model is equivalent to adding additional terms to the estimation which are usually to be small, and thus do not make a big difference.

Let us consider the generic third-order generator model:
\begin{eqnarray}
  \dot{\delta} &=& \omega\\
  M\dot{{\omega}}&=&{P_m}-{P_e}-{D}{\omega}+{\Sigma}{\xi}\\
  \dot{z}&=&f(\delta,z)
\end{eqnarray}
then the estimated Jacobian matrix can be represented as an expansion of power series in $C_{\bm{{\delta}}{\bm{{z}}}}$:
\small{
\begin{eqnarray}
  {(\frac{\partial\bm{P_e}}{\partial\bm{{\delta}}})^\star}&=&MC_{\bm{{\omega}}{\bm{{\omega}}}}C^{-1}_{\bm{{\delta}}{\bm{{\delta}}}}
  -\frac{1}{2}{C_{\bm{{\omega}}{\bm{{z}}}}C_{\bm{{\delta}}{\bm{{z}}}}}{(C_{\bm{{\delta}}{\bm{{\delta}}}}C^2_{\bm{{\omega}}{\bm{{z}}}}-C_{\bm{{\delta}}{\bm{{\delta}}}}C_{\bm{{\omega}}{\bm{{\omega}}}}C_{\bm{{z}}{\bm{{z}}}})}^{-1}\nonumber\\
  &&+O(C^2_{\bm{{\delta}}{\bm{{z}}}})
\end{eqnarray}
}
\normalsize
where $C_{\bm{{\delta}}{\bm{{z}}}}$ is typically to be small. Hence, higher-order terms can be neglected and the estimation ${(\frac{\partial\bm{P_e}}{\partial\bm{{\delta}}})^\star}=MC_{\bm{{\omega}}{\bm{{\omega}}}}C^{-1}_{\bm{{\delta}}{\bm{{\delta}}}}$ is still fairly accurate.

For numerical illustration, we consider the 9-bus 3-generator test system. %modified based on the test file ``d\_009\_tg\_mdl'' in PSAT-2.1.8,
%which is available..
Simulation was done in PSAT-2.1.8 \cite{Milano:PSAT}. All the generators are third-order models, each of which is also controlled by automatic voltage regulator. %Let the third generator be the reference, and compute
Suppose $C_{{\bm{\delta}}{\bm{\delta}}}$ and $C_{{\bm{\omega}}{\bm{\omega}}}$ can be obtained from the 100s PMUs measurements, then the estimated dynamic state Jacobian matrix is:
\begin{equation}
{(\frac{\partial\bm{P_e}}{\partial\bm{{\delta}}})^{\star}}=\left[ \begin{array}{cc} 2.127 & -0.750\\-0.861 & 1.736 \end{array}\right]
\end{equation}
and the Jacobian matrix of the deterministic model is:
\begin{equation}
{(\frac{\partial\bm{P_e}}{\partial\bm{{\delta}}})}=\left[ \begin{array}{cc} 2.150 & -0.697\\-0.708 & 1.666 \end{array}\right]
\end{equation}
Thus the estimation error is :
\begin{equation}
\frac{\|(\frac{\partial\bm{P_e}}{\partial\bm{\tilde{\delta}}})^\star-(\frac{\partial\bm{P_e}}{\partial\bm{\tilde{\delta}}})\|_F}{\|(\frac{\partial\bm{P_e}}{\partial\bm{\tilde{\delta}}})\|_F}=6.13\% \end{equation}

\end{document}